\documentclass{article}
\usepackage[letterpaper,marginparwidth=1.75cm]{geometry}

\usepackage{comment}        %
\usepackage{upgreek}
\usepackage{graphicx}        
\usepackage{multicol}        

\usepackage[colorlinks=true, allcolors=blue]{hyperref}

\usepackage{amsfonts}       
\usepackage{amsmath, amsthm}

\newcommand\independent{\protect\mathpalette{\protect\independenT}{\perp}}
\def\independenT#1#2{\mathrel{\rlap{$#1#2$}\mkern2mu{#1#2}}}
\newcommand{\R}{\mathbb{R}} 
\newcommand{\C}{\mathbb{C}} 
\newcommand{\N}{\mathbb{N}} 
\newcommand{\PP}{\mathbb{P}} 

\newcommand{\fse}{\overset{a.s.}{\longrightarrow}}
    \def\cd{\stackrel{\mathcal{D}}{\longrightarrow}}
    \def\cp{\stackrel{\mathcal{\PP}}{\longrightarrow}}

\usepackage{hyperref,url}
\newtheorem{theorem}{Theorem}
\newtheorem{lemma}{Lemma}

\newtheorem{definition}{Definition}
\newtheorem{proposition}{Proposition}
\newtheorem{remark}{Remark}
\newtheorem{example}{Example}

 \newcounter{Example}
\makeindex             

\begin{document}

\title{Independent  additive  weighted  bias distributions and associated goodness-of-fit tests}
\author{Bruno Ebner\footnote{Institute of Stochastics,
Karlsruhe Institute of Technology (KIT),
Englerstr. 2, D-76133 Karlsruhe E-mail: {bruno.ebner@kit.edu}} and Yvik Swan\footnote{Universit\'{e} libre de Bruxelles, 
D\'epartement de Mathématique -- CP 210
 Boulevard du Triomphe
B-1050 Bruxelles 
E-mail: {yvik.swan@ulb.be}}}
 
%
%
\maketitle


\abstract{We use a Stein identity  to define  a new class of parametric distributions which we call ``independent additive weighted bias distributions.''  We investigate related  $L^2$-type discrepancy measures, empirical versions of which not only encompass traditional ODE-based procedures but also offer novel methods for conducting goodness-of-fit tests in composite hypothesis testing problems. We determine critical values for these new procedures using a parametric bootstrap approach and evaluate their power through Monte Carlo simulations. As an illustration, we apply these procedures to examine the compatibility of two real data sets with a compound Poisson Gamma distribution.}

\


\section{Introduction}\label{sec:Intro}
Characterisations of distributions play a crucial role both in probability theory and in statistics. A famous example in probability theory is Stein's method, where characterisations of distributions depending on so-called Stein operators are successfully applied to distributional approximation in the sense of integral probability metrics. In  statistics, characterisations of distributions are widely used to propose goodness-of-fit and symmetry tests. The idea of exploiting characterisations for testing procedures can be traced back to \cite{Linnik1953} and became widely popular in the nineties of the last century. As noted in \cite{Nikitin2017}, goodness-of-fit tests based on characterisations are usually powerful procedures, \textit{``[...] because they reflect some intrinsic and hidden properties of probability distributions connected with the given characterisation, and therefore can be more efficient or more robust than others.''} For a historical survey on the topic see \cite{Nikitin2017}. 
The objective of this paper is to combine the characterisations of distributions utilised in Stein's method with their applications in goodness-of-fit testing.

Let $\mathcal{L}(X)$ denote the distribution of some random variable $X$. The very first step in any application of Stein's method for $\mathcal{L}(X)$ consists in identifying a linear operator $\mathcal A$ and a  wide class of functions $\mathcal F$ such that $\mathcal{L}(Y) = \mathcal{L}(X)$ if and only if $\mathbb{E}[\mathcal Af(Y)] = 0$  for all $f \in \mathcal F$. This is called a Stein characterisation, and the operator $\mathcal A$ is then called a Stein operator. The study of these operators and of their applications towards distributional approximation has attracted considerable attention, see the surveys \cite{ross, ley2016} or the monographs \cite{nourdin2012normal,shao2010stein} for an overview. Specific Stein  characterisations have, in recent years, also been  exploited in computational statistics (see \cite{anastasiou2021stein} and the many references within for an overview); in particularly the corresponding identities have already successfully been exploited  in the context of goodness-of-fit tests (see e.g. \cite{Betsch2019, betsch2020testing}).  
In this paper we propose to define an entire class of families of distributions directly through their Stein operators, then exploit this characterisation for the purpose of goodness-of-fit tests.

 The paper is structured as follows. In Section \ref{sec:some-id-distr} we define the class of families of distributions and give specific examples of parametric families contained satisfying the definition. In Section~\ref{sec:new-discr-metr} we consider a weighted $L^2$-type discrepancy measure by considering the spectral version of the Stein characterisation. Section \ref{sec:gof_test} consists of studying empirical counterparts to the discrepancy measures. We apply these for proposing new goodness-of-fit testing procedures to test composite hypotheses. In Section \ref{sec:Simu} we provide simulation results and apply in Section \ref{sec:realdata} the tests to two data sets related to insurance cases and rainfall. 


\section{A new class of families of distributions}
\label{sec:some-id-distr}

\begin{definition}[Independent Additive Weighted Bias distributions] \label{def:iawd}
A random variable  $X$ (or its distribution $\mathcal{L}(X)$) is  of \emph{independent additive weighted bias}-type if there exist
functions $a, c, d : \R  \to \R$
 and a probability distribution $\nu$ on $\R$ such
that, for $Y \sim \nu$ independent of $X$, 
\begin{equation}
  \label{eq:6}
\mathbb{E}[a(X) f(X)  - c(X) d(Y) f(X+Y)] =0
\end{equation}
for all absolutely continuous test functions $f : \R \to \R$  with polynomial growth.
We write $X \sim \mathrm{IAWD}(a, c, d, \nu)$ when (i)  $X$ satisfies \eqref{eq:6} for the prescribed functions $a, c, d$ and $Y \sim \nu$ and (ii) $\mathcal{L}(X)$  is
characterised by this identity, in the sense that if $Y$ satisfies
\eqref{eq:6} over the same class of functions then
$\mathcal{L}(Y) = \mathcal{L}(X)$.  
\end{definition}

 In the framework of Stein's method, an $\mathrm{IAWD}(a, c, d, \nu)$ random variable is characterised by an \emph{integral} Stein operator of the form $\mathcal Af(x) = a(x) f(x) - c(x) \mathbb{E}[d(Y) f(x+Y)]$. This form of operator finds its origins in the so-called size bias distributions, as follows. 

\begin{example}[Additive size bias: $a(x) = x; c(x) d(y) = c$ is 
  constant]  \label{ex:add-sibia} A positive random variable $X$
  with mean $\mu$ is said to satisfy an additive size-bias condition
  if
 $  \mathbb{E} \left[ X f(X)  - c f(X+Y)
  \right] = 0$
for some $c \in \R$ and $Y \ge 0$ independent of $X$.  Such random
variables obviously satisfy a version of \eqref{eq:6} with the
prescribed parameters; they form a subclass of the family of
infinitely divisible distributions, see \cite{arras2019stein}.  See see e.g. \cite{AGK:2019} for an overview.   The
following examples are classical: 
  \begin{enumerate}
 \item $X\sim \mathrm{Po}(\lambda)$ a Poisson random variable  with probability mass function
  $p_{\infty}(x)  \propto \lambda^x/x!$ on $\mathbb{N}$ for some
  $\lambda>0$. Then
$ \mathbb{E} \left[ X f(X) -
\lambda      f(X+Y) \right] = 0$
with $Y = 1$ almost surely.

\item $X\sim \Gamma(\alpha, \beta)$  a gamma random variable with density
  $p_{\infty}(x) \propto x^{\alpha-1}e^{-\beta x}$
  on $(0, +\infty)$ for some $\alpha, \beta>0$.  Then
$   \mathbb{E} \left[  X  f(X) - {\alpha}/{\beta} f(X +Y)\right] = 0$
  with $ X
    \independent Y \sim \mathrm{exp}(\beta)$.



\item $X \sim \mathrm{Dick}(\theta)$ a (generalized) Dickman random variable  with log-characteristic function
  $\log(\varphi_\infty(t)) =  \theta \int_0^1(e^{itu}-1)/u
    \mathrm{d}u $ for some  $\theta>0$. Then \linebreak
$  \mathbb{E} \left[ X f(X) -
     \theta f(X+Y)\right] = 0$
  where $X \independent Y \sim \mathrm{Unif}[0,1]$.


\end{enumerate}
In each case it is easy to show that these identities are characterising,  e.g. by using the test functions $f(x) = e^{i t x}$ which then leads to an ODE on the characteristic function. We will return to this in Lemma \ref{lem:tw}. 

\end{example}

Many families of distributions are  of IAWD-type,  including the  binomial, negative binomial,  and  hypergeometric distributions. In particular compound   distributions are also of IAWD-type. 

\begin{example}[Compound Poisson:  $a(x) = x; c(x) d(y) = \lambda y$]
\label{ex:compopoi}
  We say
  $X \sim \mathrm{CP}(\lambda, \nu)$ is a compound
  Poisson random variable if $X = \sum_{j=1}^N Y_i$ with  $N \sim \mathrm{Po}(\lambda)$ for some
  $\lambda>0$ independent of $(Y_i)_{i \ge 1}$ a sequence of iid
  random variables with distribution $\nu$ a probability measure on
  $\R$. Then (see \cite{arras2019stein}) its distribution is characterised by 
$ \mathbb{E} \left[ X  f(X) -
    \lambda   Y   f(X+Y)  \right] =
  0$
with $ X \independent Y \sim \nu$.
\end{example}

\begin{example} [Compound
    geometric:  $a(x) = 1; c(x) d(y) = q$]\label{ex:compogeo} We say
  $X \sim \mathrm{CGeom}(q, \nu)$ is a compound
  geometric random variable if $X = \sum_{j=1}^N Y_i$ with  $N \sim \mathrm{Geom}(p)$ for some
  $p \in (0, 1)$ independent of $(Y_i)_{i \ge 1}$ a sequence of iid
  random variables with distribution $\nu$ a probability measure on
  $\R$. Then (see \cite{daly08})
$ \mathbb{E} \left[  f(X) -
 q  f(X+Y)  \right] =
  0$
with $ X \independent Y \sim \nu$ and $q = 1-p$.
\end{example}

 Random variables  of IAWD-type have certain properties directly inherited from the identity \eqref{eq:6}. For instance, 
   \begin{enumerate}
      \item Additive size bias (Example \ref{ex:add-sibia}): if $c(x) d(y) = c$, then  $\mathbb{E}[X] = c$,  $\mathrm{Var}[X] =  c \mathbb{E}[Y]$  and $\varphi_X(t)  =  \mathrm{exp} \big( - i c \int_0^t \varphi_Y(u) \mathrm{d}u \big) $.
      \item  Compound Poisson (Example \ref{ex:compopoi}): if $c(x) d(y) = \lambda y$ with $\lambda >0$ then $\mathbb{E}[X] = \lambda \mathbb{E}[Y]$, $\mathrm{Var}[X] =  \lambda  \mathbb{E}[Y^2]$ and $\varphi_X(t)  =  \mathrm{exp} \big( -  \lambda (\varphi_Y(t) -1)\big) $.
  \end{enumerate}
More generally, relations between low order moments of $X$ and $Y$ are particularly easy to obtain through well chosen functions $f$ (typically $f(x) = 1$ and $f(x) = x$). Note moreover how   identity \eqref{eq:6} also leads to  recurrence relations on the moments of $X$ in terms of the moments of $Y$, through 
\begin{equation*}
    \mathbb{E}[a(X) X^k - c(X) d(Y) (X+Y)^k] =0
\end{equation*}
which is true for all $k$ such that $ \mathbb{E}[|a(X) X^k|]<\infty$.   Throughout this paper we will focus exclusively on examples satisfying  $a(x) = x^j$ for $j = 0, 1$ and $c(x) d(y) = c y^k$ for $k=0, 1$; for such simple functions the relations are  easy to obtain explicitly on a case-by-case basis. This will be of use in Section~\ref{sn:theteststat} for the purpose of obtaining moment estimators for the parameters of  compound distributions.

\begin{remark}
    Our definition of the IAWD family contains  an implicit assumption on the functions $a, c, d$ and on the distribution $\nu$ to ensure that identity \eqref{eq:6} is characterising. Such will be the case in all examples we consider. 
\end{remark}

\section{A new discrepancy measure}
\label{sec:new-discr-metr}

Fix some functions   $a, c, d$, some distribution $\nu$ on $\mathbb R$ and let $Y \sim \nu$.  Suppose that $X_{\infty} \sim \mathrm{IAWD}(a, c, d, \nu)$ (and in 
particular is thus characterised by its Stein identity). The subscript $\infty$ serves to emphasise that the distribution of $X_\infty$ is the target distribution in our procedure. The independent and additive structure of the biasing term in \eqref{eq:6} encourages us to consider exponential functions $f(x) = e^{\alpha x}$ in the identity; depending on whether $\alpha$ is real or complex, the Stein identity evaluated on these test functions then  leads to relations between Laplace and/or Fourier transforms of $X$ and $Y$.

In this paper we focus on the spectral  (Fourier) version:  replacing
$f(x)$ by $e^{itx}$ in \eqref{eq:6}   leads to 
\begin{align*}
D_t(X, X_{\infty}) & =  \mathbb{E} \left[ \left(
  a(X) -  c(X) d(Y)e^{itY}\right) e^{it
                     X }\right]
\end{align*}
 and, for $\omega$ some
well chosen weighting function,
\begin{align}
  \label{eq:4}
  T_{\omega}(X, X_{\infty})  = \int_{-\infty}^{\infty}   |D_{t}(X,
  X_{\infty})|^2 \omega(t) \mathrm{d}t
\end{align}
as a measure of discrepancy between $\mathcal{L}(X)$ and
$\mathcal{L}(X_{\infty})$.  
We obtain  the
following result (see the appendix for details).
 \begin{proposition}\label{prop:tdex}
   Let $\omega : \R \to \R$ be any positive, integrable and differentiable weight function. Let $X_{\infty} \sim \mathrm{IAWD}(a, c, d, \nu)$   for some $a, c, d$ and $Y \sim \nu$. We write $\mathfrak d(t) = \mathbb{E}[d(Y) e^{i t Y}]$  and  introduce the functions $\Psi_1(r) = \int_{-\infty}^{\infty} \cos(t r) \omega(t) \mathrm{d}t$, $\Psi_2(r) = \int_{-\infty}^{\infty} |\mathfrak d(t)|^2\cos(t r) \omega(t) \mathrm{d}t,$
  and $
   \Psi_3(r) = \mathbb{E}[d(Y) \Psi_1(r-Y)].
  $ Then
\begin{align}\label{eq:TXXinfomv1}
   T_\omega(X, X_{\infty})
&  = \mathbb{E} \left[   a(X_1) a(X_2)
 \Psi_1(X_1 - X_2)
      + c(X_1) c(X_2)
\Psi_2(X_1 -X_2)\right.\\
&\hspace{0.75cm}\left.- 2 a(X_1) c(X_2) \Psi_3(X_1 - X_2)\right]\nonumber
 \end{align}
 where, in \eqref{eq:TXXinfomv1}, the indexed random variables  denote
 iid copies of $X$.
\end{proposition}

\begin{remark}\label{rem:weithting}
     Many choices of weighting function $\omega$ lead to explicit and manageable expressions for the $\Psi_j$, $j=1,2, 3$. For instance, taking
  $\omega(t) \propto \mathrm{exp}({-\gamma t^2/2})$
   one easily sees that $ \Psi_1(r)=
   \mathrm{exp}(-r^2/(2\gamma))$ for all values of $r$ and all $\gamma> 0$.
  This is  obviously not the only choice. 
\end{remark}

The following holds (see the supplementary material for a proof).

\begin{lemma} \label{lem:tw}  Let $X_{\infty} \sim \mathrm{IAWB}(a, c, d, \nu)$.  Let
  $\omega$ be a weighting function which is strictly positive and integrable on $\R$, and let $T_{\omega}$ be as in
   \eqref{eq:4}.    Then  for $X$ any real random variable,
 $T_{\omega}(X, X_{\infty}) \ge 0$ with equality if and only if
   $\mathcal{L}(X) = \mathcal{L}(X_\infty)$. 
 \end{lemma}
We now detail some examples.
\begin{example}[Additive size bias  $a(x) = x; c(x) d(y) = c$]
  \label{ex:constaddsizebias} It follows   from \eqref{eq:6} (applied with $f(x) = x$) that if $X_\infty$ has finite mean, then necessarily $c = \mu$ is the mean of the law of
  $X_{\infty}$ and $\mathfrak d(t) = \mathbb{E}[e^{i t Y}]=\varphi_Y(t)$ is the characteristic function of $Y    $.  Identity \eqref{eq:TXXinfomv1}
   becomes
   \begin{align*}
           T_{\omega}(X, X_{\infty}) & =  \mathbb{E} \left[ X_1X_2 \Psi_1(X_1-X_2)
 +\mu^2 \Psi_2(X_1-X_2) - 2 \mu X_1  \Psi_3(X_1-X_2)
    \right].
   \end{align*}
  We further detail several examples
  which will be studied in the simulations (computations are left to the reader).
  \begin{enumerate}
 \item \label{it:1}
  If $X_{\infty}$ is Poisson distributed with mean $\lambda$ then  $c = \mu =  \lambda$, $Y = 1$ so that $\Psi_2(r) = \Psi_1(r)$ and
 using $\omega(t) \propto \mathrm{exp}({-\gamma t^2/2})$ as in Remark \ref{rem:weithting} we get (with slight abuse of notation)
\begin{equation}
    \Psi_1(r) = \Psi_2(r) = e^{-r^2/(2\gamma)}, \mbox{ and } \Psi_3(r) = e^{-(r-1)^2/(2\gamma)}.  \label{psipoi}
\end{equation}  \item \label{it:2}
  If $X_{\infty}$ is Dickman distributed  with mean $\lambda$ then
  $Y \sim \mathrm{Unif}[0, 1]$   for which
  $\varphi_Y(t) = (e^{i t}-1)/(i t) = \sin t / t + i (1- \cos t)/t$  so $| \mathfrak d(t)|^2 = 2(1-\cos t)/t^2 $.  Using $\omega(t) \propto t^2\mathrm{exp}({-\gamma t^2/2})$ we get
  \begin{align}
      & \Psi_1(r) = \frac{\gamma - r^2}{\gamma} e^{-r^2/(2\gamma)}, \quad   \Psi_2(r) = 2 \gamma e^{-r^2/(2\gamma)} - e^{-(r+1)^2/(2\gamma)} \left(1 + e^{2r/\gamma}\right)\gamma, \nonumber \\ 
      & \mbox{ and } \label{psidick} 
 \Psi_3(r) =   e^{-r^2/(2\gamma)}r   -  e^{-(r-1)^2/(2\gamma)} (r-1).
  \end{align}

  \item \label{it:3} If $X_{\infty}$ is gamma distributed  with parameters $\alpha, \beta$ then $c = \mu = \alpha/\beta$, $Y \sim
    \mathrm{exp}(\beta)$  for which $\varphi_Y(t) = \beta/(\beta - it)$ so  $|\mathfrak d(t)|^2 = \beta^2/(t^2 + \beta^2)$. Using $\omega(t) \propto (\beta^2+t^2)\mathrm{exp}({-\gamma t^2/2})$ we get
  \begin{align}
  & \Psi_1(r)  = \frac{\gamma - r^2+\beta^2 \gamma^2}{\gamma^2} e^{-r^2/(2\gamma)}, \quad   \Psi_2(r) = \beta^2 e^{-r^2/(2\gamma)},
 \nonumber \\
& \mbox{ and }  \Psi_3(r) = \frac{\beta}{\gamma}(r+ \beta \gamma) e^{-r^2/(2\gamma)}.
  \label{psigamma}\end{align}
    \end{enumerate}
\end{example}

\begin{example}[Compound Poisson $a(x) = x; c(x) d(y) = \lambda y$]\label{ex:compoupoi}
\hspace{-1.4mm} After recalling that $\lambda = \mu/
\mathbb{E}[Y]$,  \eqref{eq:TXXinfomv1} becomes
  \begin{align*}
    T_\omega(X, X_{\infty})
    &  = \mathbb{E} \left[   X_1X_2 \Psi_1(X_1-X_2) +  \lambda^2 \Psi_2(X_1-X_2) - 2 \lambda X_1 \Psi_3(X_1-X_2) \right].
  \end{align*}
  Note that  here  $\mathfrak d(t) = \mathbb{E}[Y e^{i t Y}]=(-i) \varphi_Y'(t)$.
   We further detail several examples
  which will be studied in the simulations (computations are left to the reader).
  \begin{enumerate}
      \item  Compound Poisson Exponential: here $Y \sim \mathrm{Exp}(\beta)$ so $|\mathfrak d(t)|^2 = \beta^2/(\beta^2+t^2)^2$.  Using $\omega(t) \propto (\beta^2+t^2)^2\mathrm{exp}({-\gamma t^2/2})$ we get
  \begin{align}\Psi_1(r) &  = \frac{1}{\gamma^4} (r^4 - 2 r^2 \gamma(3+\beta^2 \gamma) + \gamma^2 (3 + \beta^2 \gamma(2+ \beta^2 \gamma)) e^{-r^2/(2\gamma)},\nonumber \\
  \Psi_2(r) &=  \beta^2  e^{-r^2/(2\gamma)}, \mbox{ and } 
   \Psi_3(r)   = \frac{\beta}{\gamma^2} \left(-\gamma+(\beta\gamma+r)^2\right) e^{-r^2/(2\gamma)}.  \label{psipoiexp}
  \end{align}
    \item Compound Poisson Gamma: here $Y \sim \mathrm{Gamma}(\alpha, \beta)$ so \linebreak$|\mathfrak d(t)|^2 = \alpha^2/\left(\beta^2(1+(t/\beta)^2)^{\alpha+1}\right)$.  Using $\omega(t) \propto \mathrm{exp}({-\gamma t^2/2})$ we get
 $\Psi_1(r)   =  e^{-r^2/(2\gamma)}$, but   $\Psi_2$ and
   $\Psi_3$ do not admit a simple expression in terms of elementary functions. These functions can nevertheless be evaluated numerically. 
  \end{enumerate}

\end{example}

Another natural choice of test functions to consider in \eqref{eq:6} is $f(x) = e^{-tx}$ hereby leading to  $E_t(X, X_{\infty}) =  \mathbb{E} \left[ \left(
  a(X) -  c(X) d(Y)e^{-tY}\right) e^{-t
                     X }\right] $
  and, for $\omega$ some
well chosen weighting function,
$ U_{\omega}(X, X_{\infty})  = \int_{-\infty}^{\infty}   |E_{t}(X,
  X_{\infty})|^2 \omega(t) \mathrm{d}t
$
as an alternative measure of discrepancy between $\mathcal{L}(X)$ and
$\mathcal{L}(X_{\infty})$. 
Easy computations  lead to the following. 
 \begin{proposition}\label{prop:tdex2}
   Let $\omega : \R \to \R$ be any positive, integrable and differentiable weight function. Let $X_{\infty} \sim \mathrm{IAWD}(a, c, d, \nu)$   for some $a, c, d$ and $Y \sim \nu$. We write $\mathfrak e(t) = \mathbb{E}[d(Y) e^{ -t Y}]$  and  introduce the functions $\Omega_1(r) = \int_{-\infty}^{\infty} e^{-t r} \omega(t) \mathrm{d}t,$ 
 $\Omega_2(r) = \int_{-\infty}^{\infty} |\mathfrak e (t)|^2 e^{-t r} \omega(t) \mathrm{d}t, $ and $
   \Omega_3(r) =  \int_{-\infty}^{\infty} e^{-t r}\mathfrak e (t)  \omega(t) \mathrm{d}t.$
      Then
\begin{align}\label{eq:TXXinfomv2}
   U_\omega(X, X_{\infty})
&  = \mathbb{E} \left[   a(X_1) a(X_2)
 \Omega_1(X_1 + X_2)
      + c(X_1) c(X_2)
\Omega_2(X_1 + X_2)\right.\\
&\hspace{0.75cm}\left.- 2 a(X_1) c(X_2) \Omega_3(X_1 +  X_2)\right]\nonumber
 \end{align}
 where, in \eqref{eq:TXXinfomv2}, the indexed random variables  denote
 iid copies of $X$.
\end{proposition}
\begin{remark}\label{rem:comm}
If $X_\infty$ is  compound Poisson gamma distributed then for
$\omega(t) = e^{-\gamma t}$ on $\mathbb R^+ $,   the  $\Omega_i$'s
are  explicit (see Section \ref{subsim:CPG} for details). As explained in Example \ref{ex:compoupoi}, the functions $\Psi_i, i = 1, 2, 3$ are not available explicitly in this case.  
\end{remark}

\section{Goodness-of-fit tests for IAWD distributions }\label{sec:gof_test}

\subsection{The test statistics}
\label{sn:theteststat}
The goodness-of-fit testing problem for a family of distributions
$\{F(\cdot,\vartheta):\,\vartheta\in\Theta\}$, where
$\Theta\subset\R^s$ is an open parameter space, $s\in\N$, is as
follows. Let $X,X_1,X_2,\ldots$ be positive independent identically
distributed (iid) random variables and denote the distribution of
$X$ by $\mathbb{P}^X$. We want to test the \textit{null hypothesis}
\begin{equation}\label{eq:H0}
H_0:\;\mathbb{P}^X\in\{F(\cdot,\vartheta):\,\vartheta\in\Theta\}
\end{equation}
against general alternatives.  In light of preceding
arguments for any distribution of IAWD-type for some $a, c, d$ and $\nu$, we propose, for suitable weight functions $\omega$, the
weighted $L^2$-type (or Cram\'er-von Mises-type) test statistic
\begin{equation*}
  \widehat{T}_\omega^n(x_1, \ldots, x_n; X_{\infty})  =
  \int_{-\infty}^{\infty}  | \widehat{D}_t^n(x_1, \ldots, x_n; X_{\infty})
  |^2 \omega(t) \mathrm{d}t
\end{equation*}
where
$  \widehat{D}_t^n(x_1, \ldots, x_n; X_{\infty}) = \frac{1}{\sqrt n}
  \sum_{j=1}^n \left(
    \hat a(x_j)
    - \hat c(x_j) \mathbb{E}[\hat d(\hat Y) e^{i t \hat Y}]\right) e^{it x_j }
$
for $t > 0$, with $\hat a, \hat c$, $\hat d$ and $\hat Y$ needing to be estimated from the data $x_1, \ldots, x_n$.  Direct computations similar to those performed for Proposition \ref{prop:tdex} show that
\begin{align}\label{eq:testa}
 \widehat{T}_\omega^n(x_1, \ldots, x_n; X_{\infty})  &   = \frac{1}{n}
   \sum_{k, \ell = 1}^n  \left( \hat a(x_k)\hat a(x_\ell)  \hat \Psi_1(d_{k\ell})  +  \hat c(x_k) \hat c(x_\ell) \hat \Psi_2(d_{k\ell})\right.
     \\ & \hspace{1.5cm}\left.-2  \hat a(x_k)  \hat c(x_\ell) \hat \Psi_3(d_{k\ell}) \right)\nonumber
\end{align}
with $ d_{k\ell} = x_k-x_\ell$ and $\hat \Psi_j$, $j = 1, 2, 3$, are as prescribed in equation \eqref{eq:TXXinfomv1}, but this time requiring some estimation of the underlying parameters. A test based on $ \widehat{T}_\omega^n$ rejects
$H_0$ for large values of the statistic.

\begin{remark}\label{rem:MD_estimation}
    For arbitrary $\vartheta\in\Theta$ let $\widehat{T}_\omega^n(x_1, \ldots, x_n; X_{\infty};\vartheta)$ be the empirical version of the discrepancy measure in \eqref{eq:testa} in dependence of the underlying model parameters $\vartheta$. A way to estimate the unknown parameters is to calculate $$\widehat{\vartheta}_n=\mbox{argmin}_{\vartheta\in\Theta}\widehat{T}_\omega^n(x_1, \ldots, x_n; X_{\infty};\vartheta).$$
    These estimators fall into the class of minimum distance estimators and the implementation will usually need numerical routines, for a similar approach see \cite{betsch2021minimum}. We focus in the following on the goodness-of-fit testing problem and hence leave the investigation of these new estimators open for further research.
\end{remark}

\begin{example}[Additive size-bias, continued] \label{ex:adsibicont}
 Recall that $c(x) d(x) = \mu$; since the nonparametric estimate of the mean is $\bar x$ the test statistic is of the form
\begin{align}\label{eq:testaaddisizebias}
 \widehat{T}_\omega^n(x_1, \ldots, x_n; X_{\infty})  &   = \frac{1}{n}
   \sum_{k, \ell = 1}^n  \left( x_k x_\ell \hat \Psi_1(d_{k\ell})  + \bar{x}^2 \hat \Psi_2(d_{k\ell})
     -2  x_k   \bar{x} \hat \Psi_3(d_{k\ell}) \right)
\end{align}
with $\hat  \Psi_i = \Psi_i$, $i = 1, 2, 3$ as given in \eqref{prop:tdex} but with some parameters possibly  needing to be estimated. For instance, if $X_\infty$ is Poisson or  Dickman  then,   using the same weights as in Example \ref{ex:constaddsizebias}, the test statistic is given by \eqref{eq:testaaddisizebias}   with $\hat  \Psi_i = \Psi_i$, $i = 1, 2, 3$ given  without any parameter estimation in \eqref{psipoi} (Poisson case) or \eqref{psidick} (Dickman case). The same story holds if $X_\infty$ is gamma, here the functions are given from \eqref{psigamma}  but the parameter $\beta$  needs to be estimated from the data;  in our simulations  we use the moment estimator $\hat \beta = \bar x/  s^2_x$.  

\end{example}

\begin{example}[Compound Poisson,
  continued] \label{ex:compopoicontinue}
  Exactly the same extension for the compound Poisson case as in the previous example holds, here with
 \begin{align*}
  \widehat{T}_\omega^n(x_1, \ldots, x_n; X_{\infty})  &   = \frac{1}{n}
   \sum_{k, \ell = 1}^n \left( x_k x_\ell \hat \Psi_1(d_{k\ell})  + \hat \lambda^2\hat \Psi_2(d_{k\ell})
     -2  \hat \lambda  x_k    \hat \Psi_3(d_{k\ell}) \right)
  \end{align*}
where (i) for the compound Poisson exponential the $\hat \Psi_i $ for $ i = 1, 2, 3$ as given in \eqref{psipoiexp} with $\beta$ needing to be estimated whereas (ii) for the compound Poisson gamma the $\hat \Psi_i $ for $ i = 1, 2, 3$  are not available explicitly but can be computed numerically, this time with parameters $\alpha$ and $\beta$ needing to be estimated. In both cases we use the moment estimators, given in the compound Poisson exponential case by 
$    \hat \beta = 2 \bar x / s^2_x
$
and, in the compound Poisson gamma case,  by 
\begin{equation} \label{ourestgamapoi}
    \hat \alpha = \frac{B_3-2 B_2}{B_2 - B_3}, \quad \hat \beta = \frac{1}{B_3 - B_2} \quad \left( \mbox{ and } \hat \lambda = \bar x \frac{1}{B_3 - 2 B_2} \right) 
\end{equation}
with $B_2   = {s^2_x}/{\bar x} $,  $B_3  = ({\overline{x^3} -  \bar{x} \overline{x^2} - 2 \bar{x}s^2_x})/{s^2_x}$, and $\overline{x^k}=n^{-1}\sum_{j=1}^nx_j^k$.
These moment estimators are obtained by solving the  system of equations obtained by applying  identity \eqref{eq:6} to  $f(x) = 1$, $f(x) = x$ in the Poisson exponential case, and to  $f(x) = 1$, $f(x) = x$  as well as  $f(x) = x^2$ in the Poisson gamma case.   
\end{example}

\subsection{Limit null distribution}
\label{sec:limit-null-distr1}
A convenient setting for asymptotics will be the separable Hilbert
space $\mathcal{H}(\omega)$ of (equivalence classes of) measurable
functions $f:\R \rightarrow \C$ satisfying the integrability condition 
$\int |f(t)|^2 \omega(t) \, {\rm d}t < \infty$. Here,
$|z|^2=z\overline{z}$, $z\in\C$, is the complex absolute value and
$\overline{z}$ denotes the complex conjugate of $z\in\C$. We add for the notation of the mean $\overline{x}$ an index $n$ whenever this might lead to confusion. The scalar
product and the norm in $\mathcal{H}(\omega)$ will be denoted by
\begin{equation*}
  \langle f,g \rangle_{\mathcal{H}(\omega)} = \int f(t)\overline{g(t)} \, \omega(t) \, {\rm d}t, \quad \|f\|_{\mathcal{H}(\omega)} = \langle f,f \rangle_{\mathcal{H}(\omega)}^{1/2}, \quad f,g \in \mathcal{H}(\omega),
\end{equation*}
respectively. In this section we assume $X_1,\ldots,X_n$ to be iid.\ copies of $X_\infty$ and $\mathbb{E} X_\infty^2<\infty$, 
as well as $X\sim P_\vartheta$ and $Y\sim Q_{\tilde{\vartheta}}$, where $\tilde{\vartheta}\subset \vartheta$ is a vector containing some (or all) elements of the unknown parameter $\vartheta$. See Theorem \ref{theo:limitnu} of the supplementary material for the limit null distribution of $\widehat T_\omega^n$ for the case of a known parameter $\vartheta$. Denote by $\widehat{\vartheta}_n$ a consistent estimator of $\vartheta$. Write $\zeta(x,t,\vartheta)=c(x) \mathfrak{d}(t)$ for $t\in\R$ 
and assume that
\begin{enumerate}
  \item[(A1)] $\zeta$ is twice differentiable w.r.t. $\vartheta\in\Theta$, and the derivatives are bounded in some neighbourhood of $\vartheta$,
  \item[(A2)] $\widehat{\vartheta}_n$ allows an asymptotic expansion
  \begin{equation*}
  \widehat{\vartheta}_n=\vartheta+\frac1n\sum_{j=1}^n\ell(x_j,\vartheta)+o_{\mathbb{P}}(1),
  \end{equation*}
  where $\ell:\,(0,\infty)\times\R^s\rightarrow\R^s$ is a function satisfying $\mathbb{E}[\ell(X_1,\vartheta)]=0$ and $\mathbb{E}\|\ell(X_1,\vartheta)\|^2<\infty$ for all $\vartheta\in\Theta$.
  \item[(A3)] All  expectations exist.
\end{enumerate}
We introduce the stochastic process
\begin{equation*}
D_t(x_1,\ldots,x_n;\widehat{\vartheta}_n)=\frac1{\sqrt{n}}\sum_{j=1}^n(x_j-\zeta(x_j,t,\widehat{\vartheta}_n))\exp(itx_j),\quad t\in\R,
\end{equation*}
and denote in the following the gradient operator w.r.t. $\vartheta$ by $\nabla_\vartheta$ and by $x^\top$ the transpose of a vector. The proof of the subsequent theorem is found in the supplementary material file.

\begin{theorem}[limit null distribution, parameters unknown]\label{thm:CLTep}
Under assumptions (A1)-(A3) we have
\begin{equation*}
\widehat T_\omega^n(X_1, \ldots, X_n; X_{\infty})  = \|D_t(X_1,\ldots,X_n;\widehat{\vartheta}_n)\|_{\mathcal{H}(\omega)}^2\stackrel{\mathcal{D}}{\rightarrow} \|
     \mathcal{W}\|^2_{\mathcal{H}(\omega)}, \mbox{ as }n \to  \infty,
\end{equation*}
where $\mathcal{W}$ is a centred Gaussian element of
   $\mathcal{H}(\omega)$ with covariance kernel
   \begin{align*}
  K_\vartheta(s, t) & = \mathbb{E} \left[ ((X_\infty-\zeta(X_\infty,s,\vartheta))(X_\infty-\overline{\zeta(X_\infty,t,\vartheta)})\exp(i(s-t)X_\infty)\right]\\ & +\mathbb{E}\left[\exp(isX_\infty)\nabla_\vartheta\zeta(X_\infty,s,\vartheta)^\top\right]\cdot\\ &
     \ldots\cdot\left[\mathbb{E} \left[\ell(X_\infty,\vartheta)\ell(X_\infty,\vartheta)^\top\right]\mathbb{E}\left[\exp(-itX_\infty)\overline{\nabla_\vartheta\zeta(X_\infty,t,\vartheta)}\right] \right.\\ & \left.
     -\mathbb{E}\left[\ell(X_\infty,\vartheta)(X_\infty-\overline{\zeta(X_\infty,t,\vartheta)})\exp(-itX_\infty)\right] \right]\\ &
     -\mathbb{E}\left[\exp(-itX_\infty)\overline{\nabla_\vartheta\zeta(X_\infty,t,\vartheta)}^\top\right]\mathbb{E}\left[\ell(X_\infty,\vartheta)(X_\infty-\zeta(X_\infty,s,\vartheta))\exp(isX_\infty)\right].
   \end{align*}
\end{theorem}
\begin{example}[Example \ref{ex:constaddsizebias} continued]\label{ex:kerneladdsizebias} Recall that $c(y) =\mu(\vartheta)$ and $d(y) = 1$. Since the function $\zeta$   does not depend on $x$ we drop the variable in the definition for this example to gain readability. In every case $\varphi(\cdot)$ denotes the characteristic function of $X_\infty$. For the three considered cases we have
 \begin{enumerate}
 \item If $X_\infty\sim\mbox{Po}(\lambda)$ is Poisson distributed with parameter $\vartheta=\lambda>0$ then $Y = 1$ and 
 $  \zeta(t,\lambda)=\lambda\exp(it)
 $
so that  $\nabla_\lambda\zeta(t,\lambda)=\exp(it),$ $t\in\R.$
 The moment and maximum likelihood estimator is given by $\widehat{\lambda}_n=\overline{x}_n$ and hence (A2) is satisfied with $\ell(x,\lambda)=x-\lambda$. Direct calculation show
 \begin{eqnarray*}
 K_\lambda(s, t) & =& -\lambda\big[\big(\lambda(\exp(i(2s-t))+\exp(i(s-2t))-\exp(2i(s-t)))\\
 &&-(\lambda+1)\exp(i(s-t))\big)\varphi(s-t) + \exp(i(s-t))\varphi(s)\overline{\varphi(t)}\big],\quad s,t\in\R.
 \end{eqnarray*}

 \item If $X_\infty\sim\mbox{Dick}(\theta)$ is Dickman distributed with parameter $\vartheta=\theta>0$. Then $Y\sim\mbox{Unif}[0,1]$ has  characteristic function $\varphi_Y(t)=(\exp(it)-1)/it,\, t\in\R\setminus\{0\},$ $\varphi_Y(0)=1$ so that $\zeta(t,\theta)=\theta\varphi_Y(t)$ and  $\nabla_\theta\zeta(t,\theta)=\varphi_Y(t)$, 
$t\in\R.$
 A suitable estimator for $\theta$ is $\widehat{\theta}_n=\overline{x}_n$, hence (A2) is satisfied with $\ell(x,\theta)=x-\theta$. The covariance kernel in Theorem \ref{thm:CLTep} then reduces to
 \begin{eqnarray*}
 K_\theta(s, t) & =&-\varphi''(s-t)+i\theta\left(\varphi_Y(s)+\varphi_Y(-t)\right)\varphi'(s-t)-\theta^2\varphi_Y(s)\varphi_Y(-t)\varphi(s-t)\\
 &&-\varphi_Y(s)\varphi(s)\left(-\varphi''(-t)+\varphi_Y(-t)\varphi(-t)+i\theta\left(1+\varphi_Y(-t)\right)\varphi'(-t)\right)\\
 &&-\varphi_Y(-t)\varphi(-t)\left(-\varphi''(s)+\theta^2\varphi_Y(s)\varphi(s)+i\theta\left(1+\varphi_Y(s)\right)\varphi'(s)\right)\\
 &&-\frac{\theta}{2}\varphi_Y(s)\varphi_Y(-t)\varphi(s)\varphi(-t),\quad s,t,\in\R,
 \end{eqnarray*}
 since $\mathbb{E} \left[\ell(X_\infty,\vartheta)\ell(X_\infty,\vartheta)^\top\right]=\mathbb{V}(X_\infty)=\theta/2$. Note that $\varphi''(0)=-\theta^2-\theta/2$, $\varphi'(0)=i\theta$ and $\varphi(0)=1$. 
\item The case $X_\infty\sim\Gamma(\alpha,\beta)$ is detailed in Example \ref{ex:kernelgam} in the supplementary material.

 \end{enumerate}
\end{example}
The following result is a direct consequence of a Taylor expansion (see the auxiliary processes in Lemma \ref{lem:asyeq} in the supplementary material) and Fatou’s lemma.
\begin{theorem}\label{thm:consep}
Under the stated assumptions, we have as $n\rightarrow\infty$
\begin{equation*}
\liminf_{n\rightarrow\infty}\frac{\widehat{T}_\omega^{n}(X_1,\ldots,X_n;X_\infty)}{n}\ge{\rm \Delta}_{\vartheta}=\int_{-\infty}^\infty\left|D_t(X,X_\infty)\right|^2\omega(t){\rm d}t=T_\omega(X,X_\infty).
\end{equation*}
\end{theorem}
Note that $T_\omega(X,X_\infty)=0$ if and only if $\mathcal{L}(X) = \mathcal{L}(X_\infty)$ by Lemma \ref{lem:tw}. This implies that the test statistic is consistent against any alternative distribution satisfying the assumptions.

\subsection{Parametric bootstrap procedure}\label{subsec:Boot}
In this subsection assume that we have a test statistic $T_n$ for testing (\ref{eq:H0}) against general alternatives, that we have a sequence of vectors of parameters $(\vartheta_n)$ with $\vartheta_n\in\Theta$ and $\lim_{n\rightarrow\infty}\vartheta_n=\vartheta_0$ and that we have shown under a triangular array $X_{n,1},\ldots,X_{n,n}$ of row-wise iid random variables with distribution $F(\cdot;\vartheta_n)$ that $T_n\cd \|Z_{\vartheta_0}\|^2_{\mathcal{H}(\omega)}$ as $n\rightarrow\infty$, where $Z_{\vartheta_0}$ is a centred Gaussian element in $\mathcal{H}(\omega)$ with known covariance kernel $K_Z$ depending on $\vartheta_0$. The distribution of $\|Z_{\vartheta_0}\|_{\mathcal{H}(\omega)}^2$ is then known to have the equivalent representation $\sum_{j=1}^\infty\lambda_j(\vartheta_0)N_j^2$, where $N_1,N_2,\ldots$ are independent, standard normally distributed random variables, and $\lambda_1(\vartheta_0),\lambda_2(\vartheta_0),\ldots$ are a decreasing series of non-zero eigenvalues of the integral operator
\begin{equation*}
\mathcal{K}:\mathcal{H}(\omega)\rightarrow\mathcal{H}(\omega),\quad f\mapsto\mathcal{K}f(\cdot)=\int_0^\infty K_Z(\cdot,t)f(t)\omega(t)\mbox{d}t.
\end{equation*}
Clearly, the covariance kernel $K_Z$ depends on the underlying limiting parameter vector $\vartheta_0$ and hence so does the operator $\mathcal{K}$. Computing the  eigenvalues $\lambda$ of the integral operator requires solving the homogeneous Fredholm integral equation of the second kind
\begin{equation}\label{eq:inteq}
\langle K_Z(x,\cdot),f\rangle_{\mathcal{H}(\omega)}=\lambda f(x).
\end{equation}
Due to the high complexity of the covariance kernel, it seems hopeless to find explicit solutions of (\ref{eq:inteq}) and hence formulae for the eigenvalues. Furthermore, since the true parameter $\vartheta_0$ is unknown in practice, the limiting null distribution cannot be used to derive critical values of the test. Let $\widehat{\vartheta}_n$ be a consistent estimator of $\vartheta_n$, hence assume that $\widehat{\vartheta}_n\cp \vartheta_0$ as $n\rightarrow\infty$. A solution to this problem is provided by a parametric bootstrap procedure as suggested in \cite{H:1996} and which is stated as follows:
\begin{enumerate}
  \item[(1)] Compute $\widehat{\vartheta}_n=\widehat{\vartheta}_n(X_1,\ldots,X_n)$.
  \item[(2)] Conditionally on $\widehat{\vartheta}_n$ simulate $B$ bootstrap samples $X_{j,1}^*,\ldots,X_{j,n}^*$, iid from $F(\cdot;{\widehat{\vartheta}_n})$, and compute $T_{n,j}^*=T_n(X_{j,1}^*,\ldots,X_{j,n}^*)$, $j=1,\ldots,B$.
  \item[(3)] Derive an empirical $(1-\upalpha)$ quantile of $c_{n,B}^*(\upalpha)$ of $T_{n,1}^*,\ldots,T_{n,B}^*$.
  \item[(4)] Reject the hypothesis (\ref{eq:H0}) at level $\upalpha$ if $T_n(X_1,\ldots,X_n)>c_{n,B}^*(\upalpha)$.
\end{enumerate}
Note that for every computation of 
 $T_{n,j}^*$ parameter estimation
has to be done separately for each $j$.  Following the notation and
methodology of \cite{H:1996} we prove that this bootstrap test has
asymptotic level $\upalpha$ as $n,B\rightarrow\infty$. Denote the
distribution function of $T_n$ under $F(\cdot;\vartheta_n)$ by
$H_n^{\vartheta_n}(t)=\mathbb{P}_{\vartheta_n}(T_n\le t),$ $t>0,$
and write $H^{\vartheta_0}(\cdot)$ for the distribution of $\|Z_{\vartheta_0}\|_{\mathcal{H}(\omega)}^2$. Note that $H^{\vartheta_0}$ is continuous and strictly increasing on $\{t>0: 0< H^{\vartheta_0}(t) < 1\}$. By the assumptions at the beginning of this section, we have that $H_n^{\vartheta_n}(t) \rightarrow H^{\vartheta_0}(t)$ holds for every $t>0$ as $n\rightarrow\infty$, so by continuity of $H^{\vartheta_0}$ we have
\begin{equation*}
\sup_{t>0}\left|H_n^{\vartheta_n}(t)-H^{\vartheta_0}(t)\right|\longrightarrow0\quad \mbox{as}\,n\rightarrow\infty.
\end{equation*}
Combining this with the consistency of the estimators $\widehat{\vartheta}_n$, we have
\begin{equation*}
\sup_{t>0}\left|H_n^{\widehat{\vartheta}_n}(t)-H^{\vartheta_0}(t)\right|\cp0\quad \mbox{as}\,n\rightarrow\infty.
\end{equation*}
Hence, with $H_{n,B}^*(t)=\frac{1}{B}\sum_{j=1}^B\mathbf{1}\{T_{n,j}^*\le t\}$ denoting the empirical distribution function of $T_{n,1}^*,\ldots,T_{n,B}^*$, we have by an identical construction as in (3.10) of \cite{H:1996}
\begin{equation*}
\sup_{t>0}\left|H_{n,B}^*(t)-H^{\vartheta_0}(t)\right|\cp0\quad \mbox{as}\,n,B\rightarrow\infty,
\end{equation*}
from which $c_{n,B}^*(\upalpha)\cp \inf\{t:H^{\vartheta_0}(t)\ge 1-\upalpha\}$ as $n,B\rightarrow\infty$ follows. This implies that if $X_1,\ldots,X_n$ is a random sample from $F_{\vartheta_0}$, we have
\begin{equation*}
\lim_{n,B\rightarrow\infty}\mathbb{P}_{\vartheta_0}(T_n(X_1,\ldots,X_n)> c_{n,B}^*(\upalpha))=\upalpha,
\end{equation*}
ensuring an asymptotic level $\upalpha$ test.

\section{Simulations}\label{sec:Simu}
We now present the results of Monte Carlo simulation studies related to the tests discussed in the previous section. Because these are the most novel tests, we focus on the examples of testing the fit to the gamma distribution, the generalised Dickman distribution, and the compound Poisson gamma law. Other examples are illustrated in the supplementary material. In each case the distributional parameters are unknown and have to be estimated. As a consequence a parametric bootstrap procedure is needed to perform the tests. For the first two families of distributions, we implemented the procedure stated in Subsection \ref{subsec:Boot} and for the last family the warp-speed method of \cite{GPW:2013} due to heavy numerical computation times. We refer to Section \ref{sec:simuSM} of the supplementary material for further simulation results on testing for the Poisson and the compound Poisson exponential distributions.

All the simulations are performed using the statistical computing environment \texttt{R}, see \cite{R:20}. For the first three cases, we consider the sample size $n=50$ and the nominal level of significance is set either to $\upalpha=0.05$ or to $\upalpha=0.1$ depending on the comparable simulation studies in the literature. Every entry in Tables \ref{tab:Poisson} - \ref{tab:CPE} in the supplementary material are based on 10 000 repetitions with a bootstrap sample of $B=500$. For the compound Poisson gamma law, we fix the simulation parameters to $n=100$, $\upalpha=0.05$, and perform 1 000 repetitions. To ensure easy readability the tables containing the simulation results and the description of the alternative distributions have been moved to the supplementary materials file.



\subsection{Testing the fit to the gamma family of distributions}\label{sssec:gamma}
The problem of testing the fit of data to the gamma distribution with unknown parameters is considered in the literature, see \cite{BE:2019, HME:2012} and the references therein. Note that most of the considered procedures are implemented in the \texttt{R}-package \texttt{gofgamma}, see \cite{BEB:20}. 
As described in Example~\ref{ex:adsibicont}, we compute the test statistic \eqref{eq:testaaddisizebias}  with functions $\hat \Psi_i, i = 1, 2, 3$ given in \eqref{psigamma} 
with scale parameter replaced with its   moment estimator. 

Firstly, we see that the type I error is correctly controlled by the bootstrap procedure, although it results in conservative procedures. A comparison of the empirical powers in Table \ref{tab:Gamma} from the supplementary material   to the results given in Table 2 of \cite{BE:2019} shows that for the SP(2) distribution the newly proposed procedure outperform the existing procedures. On many other cases there are choices of the tuning parameter $\gamma$ for which the tests are close to the best performing competitors. Clearly some alternatives as the W(1.5) law are not identified at all.

\subsection{Testing the fit to the family of generalized Dickman distributions}
The Dickman (or alternatively Dickman-Goncharov) distribution is a distribution appearing as a limiting distributions connected to the analysis of the asymptotic behaviour of the number of positive integers in an interval, where the largest prime factor is smaller than a functional of the length of the interval. For a historic overview and further applications see \cite{MP:2020}. 
The problem of testing the fit of data to a generalized Dickman distribution (which follows the definition in \cite{arras2019stein}) has not yet been considered in the literature and hence our test statistic presented in Example \ref{ex:adsibicont} (obtained from \eqref{eq:testaaddisizebias}  with functions $\hat \Psi_i, i = 1, 2, 3$ given in \eqref{psidick})  is the only test for this family of distributions. Note that the classical procedures based on the empirical distribution function can't be computed, since there exists no closed form formula for the distribution function of this law. As estimators of the unknown parameter we chose the moment estimator. Since the support of the Dickman distribution is $\R_+$ we choose the notation for the alternative families of distributions given in  Section \ref{sssec:gamma}. Random number generation for the generalized Dickman law was performed using the Algorithm 3.1 of \cite{DQL:2019} implemented in the \texttt{R} package \texttt{SubTS}, see \cite{GC:2023}.

The power estimates given in Table \ref{tab:Dickman} in the supplementary materials show that the type I error is controlled for the significance level $\alpha=0.1$ under the hypothesis and the alternative distributions are well identified.


\subsection{Testing the fit to the family of compound Poisson gamma distributions}\label{subsim:CPG}
In this subsection we want to test the fit of data to the compound Poisson gamma distribution, see Example \ref{ex:compopoi} where $\nu$ is the two parameter gamma distribution $\Gamma(\alpha,\beta)$, $\alpha,\beta>0$. This family of distributions was used to model aggregated insurance claims \cite{HTD:2022,XG:2015}, weighted networks in finance \cite{GV:2020}. As estimators of the three unknown parameters $(\lambda,\alpha,\beta)$ we used the method of moments estimators \eqref{ourestgamapoi} and denote them by $(\widehat{\lambda}_n,\widehat{\alpha}_n,\widehat{\beta}_n)$ respectively. 

As already commented upon in example \ref{ex:compoupoi}, for compound Poisson gamma distributions  the test statistic $\widehat T_\omega$  from \eqref{eq:TXXinfomv1} 
 suffers  the drawback that the coefficients $\widehat \Psi_i$, $i=1, 2, 3$ do not admit an explicit form and therefore need to be computed numerically. On the other hand,   $U_\omega$ from \eqref{eq:TXXinfomv2}  with weight function  $\omega(t) = \mathfrak{e}^{-1}(t) e^{-\gamma t}$ on $(0, \infty)$  for $\gamma>0$ 
 a tuning parameter admits explicit coefficients, leading to 
\begin{equation*}
    \widehat U_\gamma=\frac1n\sum_{j,k=1}^n x_jx_kK_\gamma^{(2)}(x_j+x_k)+\widehat{\lambda}_nx_j K_\gamma^{(1)}(x_j+x_k)+\widehat{\lambda}_n^2(x_j+x_k+\gamma)^{-1},
\end{equation*}
where 
\begin{eqnarray*}
    K_\gamma^{(1)}(x)&=&-\frac{\exp\left((x+\gamma)/\widehat{\beta}_n\right)\widehat{\beta}_n^{\widehat{\alpha}_n}}{\widehat{\alpha}_n(x+\gamma)^{\widehat{\alpha}_n+2}}\Gamma_{iu}\left(\widehat{\alpha}_n+2;(x+\gamma)/\widehat{\beta}_n\right),\\
    K_\gamma^{(2)}(x)&=&\frac{\exp\left((x+\gamma)/\widehat{\beta}_n\right)\widehat{\beta}_n^{2\widehat{\alpha}_n}}{\widehat{\alpha}_n^2(x+\gamma)^{2\widehat{\alpha}_n+3}}\Gamma_{iu}\left(2\widehat{\alpha}_n+3;(x+\gamma)/\widehat{\beta}_n\right),
\end{eqnarray*}
and $\Gamma_{iu}(\alpha,x)=\int_x^\infty y^{\alpha-1}\exp(-y)\,\mbox{d}y$ denotes the upper incomplete gamma function. 
As already commented upon in Remark \ref{rem:comm},  this is  precisely the test statistic proposed by \cite[Section 3.2]{GJM:2022} (note that there is a typographical error in the stated formula in display (35)) for  the particular case of CPG distributions. 

We consider the alternative distributions from Subsection \ref{sssec:gamma} described in Subsection \ref{subsec:GDSM} of the supplementary material and add a mixed compound Poisson gamma model by simulating MCP$({\rm p};\lambda_1,\Gamma(\alpha_1,\beta_1),\lambda_2,\Gamma(\alpha_2,\beta_2))={\rm p}\,\mbox{CP}(\lambda_1,\Gamma(\alpha_1,\beta_1))+(1-{\rm p})\mbox{CP}(\lambda_2,\Gamma(\alpha_2,\beta_2))$ for $\lambda_1=\alpha_1=1$, $\beta_1=\beta_2=3$,  $\lambda_2\in\{1,2,3,4,5,10,20,50\}$, $\alpha_2\in\{5,6\}$, and ${\rm p}\in\{1/4,1/2,3/4\}$. Since the simulation was numerically involved we restricted the study to the tuning parameter $\gamma=1$. The results are shown in Table~\ref{tab:CPG}. We see that both methods tend to be conservative although for the CP$(1,\Gamma(1,5))$ law they both tend to overestimate slightly the nominal level of $0.05$. This effect might be explained by the low amount of Monte Carlo repetitions. In most cases the Laplace transform method outperforms the  test based on $\widehat{T}_1$ using the weight function $\omega(t)\propto \exp(-\gamma t^2/2)$ with tuning parameter $\gamma=1$, although for Weibull and some mixtures of compound Poisson gamma distributions the power is better in the reverse order. These findings are not surprising, since there is no universally best goodness-of-fit test, see \cite{janssen2000global}. 

\begin{table}[t]
\centering
\caption{Empirical rejection rates for the tests of compound Poisson gamma distribution ($n=100$, $\upalpha=0.05$, 1000 repetitions, warp speed bootstrap)}\label{tab:CPG}
\begin{tabular}{lrr}
Dist./Test & $\widehat{T}_1$ & $\widehat U_1$ \\ 
  \hline
CP$(1,\Gamma(1,5))$ & 6 & 8 \\ 
CP$(2,\Gamma(1/2,5))$ & 3 & 4 \\ 
CP$(3,\Gamma(1/3,5))$ & 1 & 5 \\ 
CP$(4,\Gamma(3/2,5))$ & 2 & 4 \\ \hline
$\Gamma(1/4)$ & 5 & 19 \\ 
$\Gamma(5)$ & 5 & 7 \\ 
$\Gamma(10)$ & 4 & 8 \\ 
IG$(1/2)$ & 15 & 61 \\ 
IG$(3/2)$ & 13 & 54 \\ 
LN(0.5)  & 8 & 22 \\ 
LN(0.8)  & 15 & 49 \\ 
W(3, 2) & 6 & 1 \\ 
W(2, 5) & 7 & 3 \\ 
MCP$(1/4,1,\Gamma(1,3),1,\Gamma(6,3))$ & 7 & 18 \\ 
MCP$(1/4,1,\Gamma(1,3),2,\Gamma(6,3))$ & 11 & 10 \\ 
MCP$(1/4,1,\Gamma(1,3),3,\Gamma(6,3))$ & 14 & 9 \\ 
MCP$(1/4,1,\Gamma(1,3),4,\Gamma(6,3))$ & 22 & 10 \\ 
MCP$(1/2,1,\Gamma(1,3),1,\Gamma(5,3))$ & 4 & 3 \\ 
MCP$(1/2,1,\Gamma(1,3),2,\Gamma(5,3))$ & 10 & 8 \\ 
MCP$(1/2,1,\Gamma(1,3),3,\Gamma(5,3))$ & 30 & 53 \\ 
MCP$(1/2,1,\Gamma(1,3),4,\Gamma(5,3))$ & 46 & 84 \\ 
MCP$(3/4,1,\Gamma(1,3),5,\Gamma(5,3))$ & 31 & 44 \\ 
MCP$(3/4,1,\Gamma(1,3),10,\Gamma(5,3))$ & 60 & 100 \\ 
\end{tabular}
\end{table}

\section{Real data application}\label{sec:realdata}
Since the presented method is very general we focus our data examples on the compound Poisson Gamma family of distributions from Example \ref{ex:compoupoi}. The two most found applications of this law found in the literature are for insurance and rainfall data.

\subsection{Insurance data}
In our first example we analyse  the data set of insurance claims of motorcycle drivers collected in 1999 by the Swedish insurance company Wasa. This data set contains information about motorcycle policies over the time period from 1994 to 1998 containing quantitative and categorical variables like 'owner age', 'gender', 'vehicle age', 'bonus class', 'number of claims' and 'claim cost'. The data is available on the companion website of the monograph \cite{OJ:2010}. This data set has been analysed in Section 4 in \cite{HTD:2022}, where assumptions are made for the quantitative variables 'number of claims' and 'claim cost' to be distributed following a Poisson and gamma law, respectively. In the following paragraphs,  we aim at studying the validity of the assumptions  by the methods  proposed  in the previous sections. Since 'owner age' and 'bonus class' are arguably good parameters to describe the experience of a driver we choose appropriate subsets of the data set. In a first step we analyse the validity of the assumption that the number of claims follow a Poisson law by the test statistic given in item 1 of Example \ref{ex:adsibicont} and applying the parametric bootstrap procedure of Subsection \ref{subsec:Boot} with 500 bootstrap samples. 
In Table \ref{tab:pois.real} we provide bootstrap p-values by relative frequency of simulated $T_{n,1}^*,\ldots,T_{n,500}^*$ that are smaller than $\widehat{T}_\omega^n(x_1,\ldots,x_n;X_\infty)$. It is seen that in most cases the assumption of an underlying Poisson distribution cannot be rejected at a significance level of $5\%$. Nevertheless there are combinations, where the assumption of Poissonity is clearly rejected, as is the case of Bonusclass 1 and Ownerage $[25,30)$. A closer look 
into this class shows that the data set has $n=1731$ values of which 21 are greater than zero, 19 of which take the value 1 and 2 take the value 2. The rest of the data are zero. All the other p-values smaller than 0.05 exhibit a similar 
or even smaller amount of cases, where the test rejects the hypothesis. This shows that in such sparse data sets the assumption of Poissonity should not be taken for granted.

\begin{table}[!t]
\caption{Bootstrap p-values poisson law (bootstrap sample size 500), 0* stands for data containing only 0's, hence no p-value could be computed}\label{tab:pois.real}
\centering \small 
\begin{tabular}{rrrrrrrrr}
  Bonusclass & & & & & & & & \\
  Ownerage & $<20$ & $[20,25)$ & $[25,30)$ & $[30,35)$ & $[35,40)$ & $[40,50)$ & $[50,60)$ & $\ge60$ \\
  \hline
1 & 0.182 & 0.030 & 0.000 & 0.004 & 0.424 & 0.486 & 0.004 & 0.328 \\
  2 & 0.538 & 0.146 & 0.548 & 0.460 & 0.522 & 0.424 & 0.454 & 0.306 \\
  3 & 0.368 & 0.484 & 0.506 & 0.438 & 0.578 & 0.510 & 0.002 & 0.546 \\
  4 & 0* & 0.214 & 0.610 & 0.574 & 0.556 & 0.474 & 0.572 & 0.356 \\
  5 & 0* & 0.064 & 0.068 & 0.004 & 0* & 0.416 & 0.352 & 0* \\
  6 & 0* & 0.432 & 0.518 & 0.406 & 0.548 & 0.442 & 0.332 & 0.534 \\
  7 & 0* & 0.580 & 0.028 & 0.068 & 0.472 & 0.002 & 0.000 & 0.484 \\
\end{tabular}
\end{table}

\begin{figure}[b]
\includegraphics[width=7.5cm]{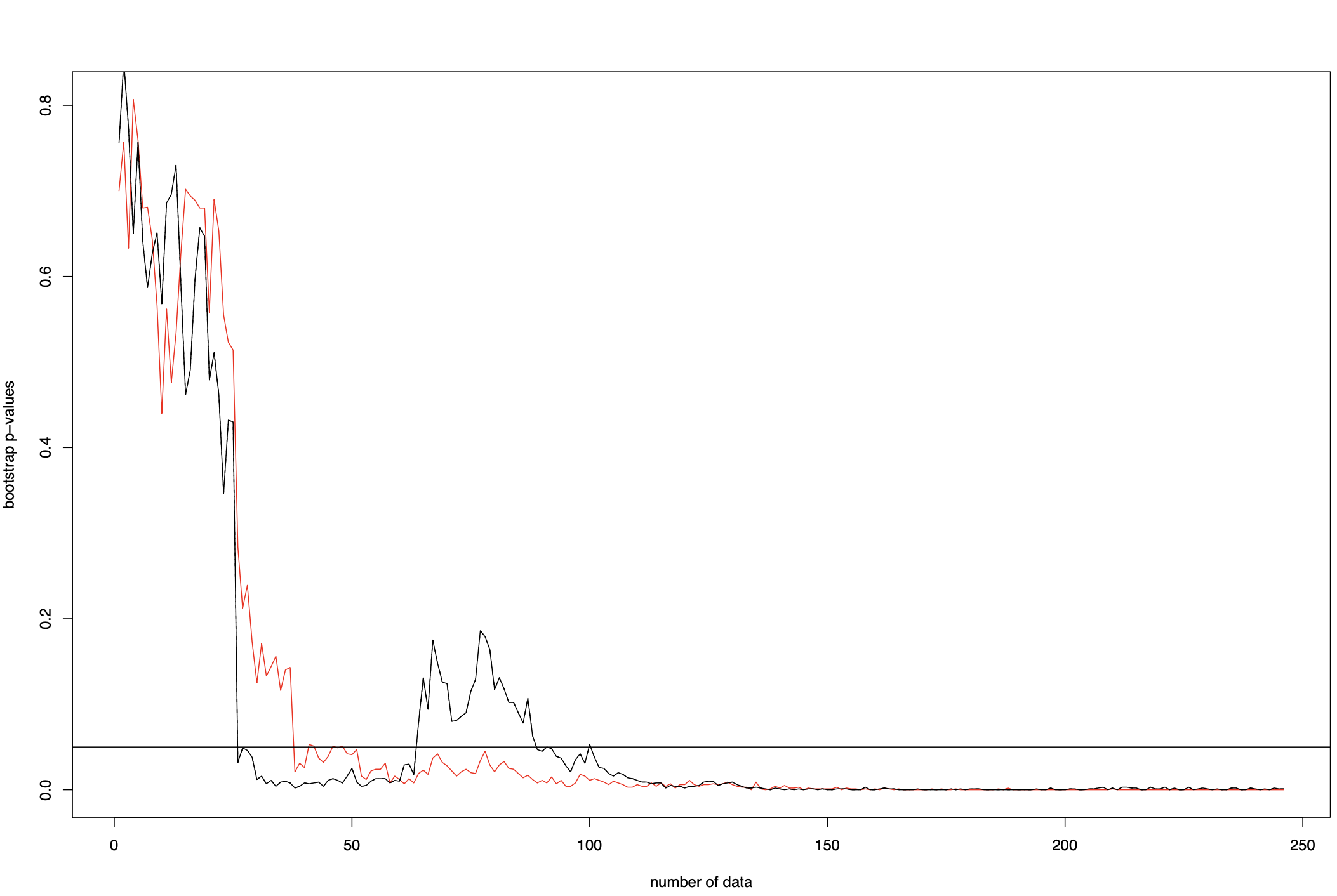}
\caption{Bootstrap p-values (bootstrap sample size 1000) for $T_1$ (red) and $T_2$ (black) with tuning parameter 1.}\label{fig:gamma.real}
\end{figure}

In a second step we consider the subset of data in which exactly one claim occured, which happened in $n=639$ cases. In Figure \ref{fig:gamma.real} we plotted the bootstrap p-value (bootstrap sample size 1000) for subsets of this data where the first 5 to 250 values have been considered. A line for a significance level of 5\% has been added to show the rejection of the assumption that claims are distributed following a gamma distribution already for small sample sizes of less than 30. Taking all data into account, both procedures return a bootstrap p-value of 0 so the hypothesis is rejected as was to be expected.

As a last step, we look at the comparably more involved case, where we limit the data to only observations of the aggregated claim cost without the knowledge of how many claims have been reported. This setting falls into the class of compound Poisson gamma distributions. Since heavy numerical integration routines have been used, we limit our study to the cases reported in Table \ref{tab:CPG.real}. The asterisk '*' stands for cases where we could not compute a p-value due to numerical problems. Interestingly the flexibility of the model seems to compensate the drawbacks reported in the first two tests. Nevertheless we also have to reject the hypothesis for the groups involving the owner ages $[20,25)$ at a 5\% level.

\begin{table}[!t]
\caption{Bootstrap p-values compound Poisson gamma (bootstrap sample size 100)}\label{tab:CPG.real}
\centering
\begin{tabular}{lrrrr}
Bonusclass$\backslash$Ownerage & $<20$ & $[20,25)$ & $[25,30)$ & $[30,35)$ \\ \hline
1 & 0.23 & 0.03 & 0.38 & 0.33\\
2 & * & 0.05 & 0.33 & *\\
3 & * & * & 0.35 & * \\
4 & * & * & 0.09 & *
\end{tabular}
\end{table}

\subsection{Rainfall data}
\begin{figure}[b]
\includegraphics[width=7.5cm]{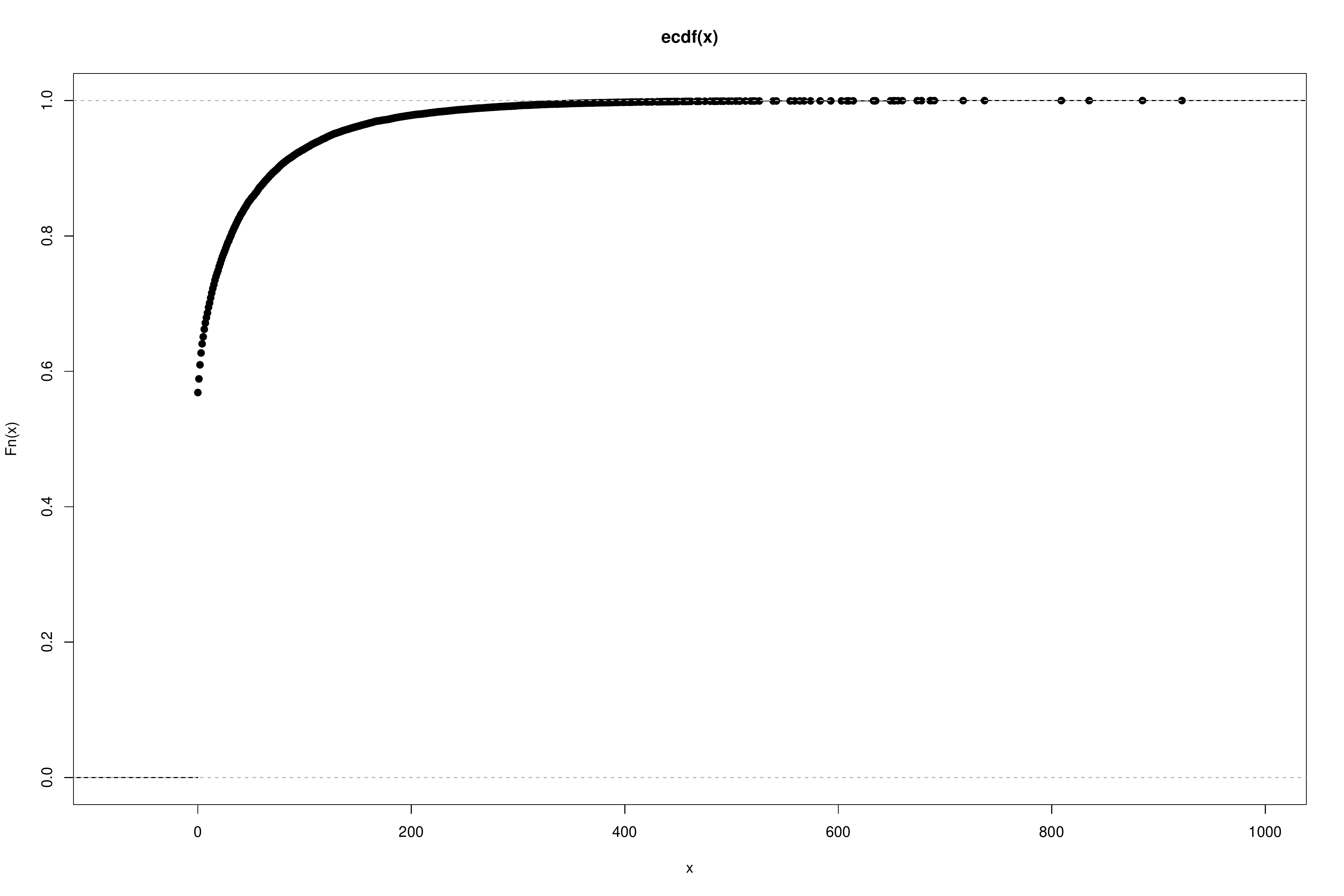}
\caption{Empirical distribution function of the Innsbruck rainfall data}\label{fig:rain.ecdf}
\end{figure}

As a second example we consider rainfall data from the European Climate Assessment \& Dataset (ECA\&D, see www.ecad.eu), see \cite{K:2002}. We consider data taken from the station ID 13, where the data is gathered at Innsbruck University in Austria. The collected precipitation amount is given (in 0.1 mm) on a daily basis, covering the time span from 1st of January 1877 to 31st of July 2022. The data set consists of 53168 entries, after removing 4 \texttt{NA} entries. At 30234 days no rainfall was recorded. A plot of the empirical distribution function is found in Figure \ref{fig:rain.ecdf}.
Since such large data sets are numerically not tractable, we restrict ourselves to the first and last year of recording, namely from 1st of January 1877 to 31st of December 1877 and 1st of August 2021 to 31st of July 2022. The values of the estimators are $(0.793804619, 0.295286094, 0.008799322)$ and $(0.35834374, 1.09216710, 0.01713137)$, respectively. The tests calculate a bootstrap p-value of 0.38 and 0.54 (bootstrap sample size 100). Hence, we cannot reject the null hypothesis that the data stems from a compound Poisson gamma distribution at any level.

\bibliographystyle{abbrv}



\section*{Acknowledgement}
We thank Julien Trufin for providing us with the database used in Section \ref{sec:realdata}. 

\appendix

\section{Proofs and additional results}\label{sec:proofs}

\begin{proof}[Proof of Lemma \ref{lem:tw}]
   Positivity and  necessity are trivial. For sufficiency, let us suppose that
   $T_{\omega}(X, X_{\infty})= 0$. Then $D_t(X, X_{\infty}) = 0$ for
   (almost) all $t$. Using the inverse Fourier formula  and  Fubini's theorem, it follows that for any continuous  integrable $f$
\begin{align*}
  \mathbb{E} \left[ a(X) f(X)  - c(X) d(Y) f(X+Y)\right] & =
                                                           \int \hat
                                                           f(\xi)
                                                           D_{2 \pi \xi}(X, X_{\infty})
                                                           \mathrm{d}\xi
                                                           = 0
\end{align*}
with  $\hat f$
   the   Fourier transform of $f$.
Hence  $X$ satisfies the same Stein identity \eqref{eq:6}  as $X_\infty$.
 \end{proof}
\begin{theorem}[Limit null distribution, parameters known] \label{theo:limitnu}
  Let  $X_{\infty}$  satisfy
  \eqref{eq:6} for some functions $a, c$ and $d$. We write
   $d_1(t) = \mathbb{E} \left[ d(Y) \cos (tY) \right]$ and
   $d_2(t) = \mathbb{E} \left[ d(Y) \sin (tY) \right]$.   Then under the assumptions of Section \ref{sec:limit-null-distr1}
  \begin{equation}
    \label{eq:5}
     \widehat T_\omega^n(X_1, \ldots, X_n; X_{\infty})  = \|  \widehat
     D^n(X_1, \ldots, X_n; X_{\infty})\|_{\mathcal{H}(\omega)}^2
     \stackrel{\mathcal{D}}{\rightarrow} \|
     \mathcal{W}\|^2_{\mathcal{H}(\omega)}, \mbox{ as }n \to  \infty,
   \end{equation}
   where $\mathcal{W}$ is a centered Gaussian element of
   $\mathcal{H}(\omega)$ with (real) covariance kernel
   \begin{align*}
     K(s, t) & = \mathbb{E} \left[ A_1(X_{\infty}, s, t)
              \cos((t-s)X_{\infty}) + A_2(X_{\infty}, s, t)
               \sin((t-s)X_{\infty})  \right]
   \end{align*}
   with
   \begin{align*}
A_1(x, s, t) & = a(x)^2 +
               a(x)c(x) (d_1(s) + d_1(t))  +    c(x)^2 \left( d_1(s) d_1(t) +
            d_2(s) d_2(t) \right) \\
A_2(x, s,t) & =
               a(x)c(x) (d_2(s) - d_2(t))  +    c(x)^2 \left( d_1(s) d_2(t) -
                                                                                d_2(s)
                                                                                d_1(t)\right)
   \end{align*}
   where $d_1(t) = \mathbb{E} \left[ d(Y) \cos(tY) \right]$ and
   $d_2(t) = \mathbb{E} \left[ d(Y) \sin(tY) \right]$.
   Under the stated assumptions, we have as $n\rightarrow\infty$
\begin{equation*}
\frac{\widehat T_\omega^n(X_1, \ldots, X_n; X_{\infty})}{n}\fse{\rm \Delta}_{\lambda_0}=\int_{-\infty}^\infty\left|D_t(X,X_\infty)\right|^2\,\omega(t){\rm d}t=T_\omega(X,X_\infty).
\end{equation*}
 \end{theorem}

 \begin{proof}[Proof of Theorem \ref{theo:limitnu}] It is direct to see that
   \begin{align*}
     K(s, t) & = \mathbb{E} \left[e^{i(t-s)X_{\infty}} \left(
               a(X_{\infty}^2) - a(X_{\infty})c(X_{\infty}) \left(
               \mathfrak d(t) + \overline{\mathfrak d(s)} \right) +
               c(X_{\infty})^2 \mathfrak d(t) \overline{\mathfrak d(s)} \right)
               \right]
   \end{align*}
   with $\mathfrak d(t) = \mathbb{E} \left[ d(Y)e^{i t Y} \right]$. Some
   trivial rearrangements then give
    \begin{align*}
     K(s, t) & = \mathbb{E} \left[ A_1(X_{\infty}, s, t)
              \cos((t-s)X_{\infty}) + A_2(X_{\infty}, s, t)
               \sin((t-s)X_{\infty})  \right] \\
     & \quad + i \mathbb{E} \left[ A_1(X_{\infty}, s, t)
              \sin((t-s)X_{\infty}) + A_2(X_{\infty}, s, t)
\cos((t-s)X_{\infty})  \right]
    \end{align*}
    with all expressions as defined in the statement of the
    Theorem. Now note that $A_1(x, s,t) = A_1(x, t, s)$ so that  $\mathbb{E} \left[ A_1(X_{\infty}, s, t)
              \sin((t-s)X_{\infty}) \right]=0$ and $A_2(x, s, t)  = -
            A_2(x, t, s)$ so that $\mathbb{E} \left[ A_2(X_{\infty}, s, t)
\cos((t-s)X_{\infty})  \right] = 0$, whence the claim.
 \end{proof}

\begin{proof}[Proof of Proposition \ref{prop:tdex}]
  Let $\mathfrak d(t) = \mathbb{E}[d(Y) e^{i t Y}] = \mathrm{Re}(\mathfrak d (t))  + i \mathrm{Im}(\mathfrak d(t)) $.  Straightforward but tedious computations (we used \texttt{Mathematica} and symmetry arguments to simplify the expressions)  lead to
  \begin{align*}
      | D_t(X, X_\infty) |^2 & =  \mathbb{E} \left[ \left(
  a(X) -  c(X) d(Y)e^{itY}\right) e^{it
                     X }\right]\mathbb{E} \left[ \left(
  a(X) -  c(X) d(Y)e^{-itY}\right) e^{-it
                     X }\right]\\
                     & = \mathbb{E} \left[ \left(
  a(X_1) -  c(X_1) \mathfrak d(t)\right) e^{it
                     X_1 } \left(
  a(X_2) -  c(X_2) \mathfrak d(-t)\right) e^{-it
                     X_2 }\right]\\
               & = \mathbb{E}\left[ \left(a(X_1) a(X_2) +|\mathfrak d(t)|^2 c(X_1 ) c(X_2) \right)\cos(t (X_1 - X_2 ))\right] \\
               & \quad - \mathbb{E} \left[ \left( a(X_1) c(X_2) + c(X_1) a(X_2)\right) \mathrm{Re}(\mathfrak d (t)) \cos(t(X_1 - X_2))\right] \\
               & \quad - \mathbb{E} \left[ \left(  a(X_1) c(X_2) - c(X_1) a(X_2)\right) \mathrm{Im}(\mathfrak d(t)) \sin(t(X_1 - X_2))\right].
  \end{align*}
  Clearly,
    \begin{align*}
  & \mathbb{E} \left[ \left( a(X_1) c(X_2) + c(X_1) a(X_2)\right) \mathrm{Re}(\mathfrak d (t)) \cos(t(X_1 - X_2))\right. \\
  & \quad +\left. \left( a(X_1) c(X_2) - c(X_1) a(X_2)\right) \mathrm{Im}(\mathfrak d (t)) \sin(t(X_1 - X_2))\right] \\
               & = \mathbb{E} \left[ d(Y) \left( \left( a(X_1) c(X_2) + c(X_1) a(X_2)\right) \cos(tY)  \cos(t(X_1 - X_2)) \right.\right. \\ & \qquad +\left. \left. \left( a(X_1) c(X_2) - c(X_1) a(X_2)\right)  \sin(tY) \sin(t(X_1 - X_2))\right)\right] \\
              & = \mathbb{E} \left[ d(Y)  \left( a(X_1) c(X_2) \cos(t(X_1 - X_2 - Y)) + c(X_1) a(X_2) \cos(t(X_1 - X_2+Y)) \right)\right]  \\
              & =  2 \mathbb{E} \left[ d(Y)  a(X_1) c(X_2) \cos(t(X_1 - X_2 - Y))\right]  .
    \end{align*}
    The conclusion follows.
\end{proof}

Before formulating the proof of Theorem \ref{thm:CLTep}, we introduce the auxiliary processes
\begin{equation*}
\widetilde{D}_t(X_1,\ldots,X_n;\vartheta)=\frac1{\sqrt{n}}\sum_{j=1}^n(X_j-\zeta(X_j,t,\vartheta))\exp(itX_j)-\exp(itX_j)\nabla_\vartheta\zeta(X_j,t,\vartheta)^\top(\widehat{\vartheta}_n-\vartheta)
\end{equation*}
as well as
\begin{equation*}
\check{D}_t(X_1,\ldots,X_n;\vartheta)=\frac1{\sqrt{n}}\sum_{j=1}^n(X_j-\zeta(X_j,t,\vartheta))\exp(itX_j)-\mathbb{E}\left[\exp(itX_1)\nabla_\vartheta\zeta(X_1,t,\vartheta)^\top\right]\ell(X_j,\vartheta),
\end{equation*}
and prove the following Lemma.
\begin{lemma}\label{lem:asyeq}
Under the standing assumptions, we have
\begin{equation*}
\|D_t-\widetilde{D}_t\|_{\mathcal{H}(\omega)}=o_\mathbb{P}(1)\quad\mbox{and}\quad\|\widetilde{D}_t-\check{D}_t\|_{\mathcal{H}(\omega)}=o_\mathbb{P}(1).
\end{equation*}
\end{lemma}

\begin{proof}[Proof of Lemma \ref{lem:asyeq}]
Denoting by $H$ the Jacobian matrix w.r.t. $\vartheta$, we have with a multivariate Taylor expansion of order 2 and by the Cauchy-Schwarz inequality 
\begin{eqnarray*}
\|D_t-\widetilde{D}_t\|^2_{\mathcal{H}(\omega)}&=&\int_{-\infty}^\infty\left|\frac1{\sqrt{n}}\sum_{j=1}^n(\widehat{\vartheta}_n-\vartheta)^\top H(X_j,t,\vartheta_0)(\widehat{\vartheta}_n-\vartheta)\exp(itX_j)\right|^2\omega(t)\mathrm{d} t\\
&\le&\|\widehat{\vartheta}_n-\vartheta\|^2\|\sqrt{n}(\widehat{\vartheta}_n-\vartheta)\|^2\frac1{n^2}\sum_{j,k=1}^n\int_{-\infty}^\infty\left\|M(X_j,t)M(X_k,t)\right\|_F^2\omega(t)\mathrm{d}t,
\end{eqnarray*}
where $\vartheta_0$ lies between $\widehat{\vartheta}_n$ and $\vartheta$, $M$ is the upper bound of $H$ w.r.t. $\vartheta$ (which exists by (A1)), and $\|\cdot\|_F$ denotes the Frobenius norm. By the sub-multiplicative structure of the latter we have by the Cauchy-Schwarz inequality 
\begin{eqnarray*}
\frac1{n^2}\sum_{j,k=1}^n\int_{-\infty}^\infty\left\|M(X_j,t)M(X_k,t)\right\|_F^2\omega(t)\mathrm{d}t\le\left(\frac1n\sum_{j=1}^n\int_{-\infty}^\infty\left\|M(X_j,t)\right\|_F^2\omega(t)\mathrm{d}t\right)^2
\end{eqnarray*}
and an application of the strong law of large numbers shows with (A2) and (A3) the claim. The second statement follows by the law of large numbers in Hilbert spaces and (A2).
\end{proof}

\begin{proof}[Proof of Theorem \ref{thm:CLTep}]
By Lemma \ref{lem:asyeq}, we see that the asymptotic distribution of $D_t$ is determined by $\check{D}_t$ and the latter is a sum of iid random variables, hence the central limit theorem in Hilbert-spaces can be applied. The statement is then a direct consequence of the continuous mapping theorem. Writing 
\begin{equation*}
    W_1(t)=(X_1-\zeta(X_1,t,\vartheta))\exp(itX_1)-\mathbb{E}\left[\exp(itX_1)\nabla_\vartheta\zeta(X_1,t,\vartheta)^\top\right]\ell(X_1,\vartheta),
\end{equation*}
the covariance structure of the limit process is given by $\mathbb{E}\left[W_1(s)\overline{W_1(t)}\right]$ and straightforward calculations.
\end{proof}
In the following we detail the covariance kernel for the gamma distribution with unknown parameters. 
\begin{example}{Example \refstepcounter{Example}\theExample\label{ex:kernelgam}: Example \ref{ex:kerneladdsizebias} continued.}
If $X_\infty\sim\Gamma(\alpha,\beta)$ is gamma distributed with parameter vector $\vartheta=(\alpha,\beta)\in(0,\infty)^2$. Then $Y\sim\mbox{Exp}(\beta)$ with characteristic function $\varphi_Y(t)=\beta/(\beta-it)$, $t\in\R$. Then
 \begin{equation*}
     \zeta(t,\vartheta)=\frac{\alpha}{\beta}\varphi_Y(t)=\frac{\alpha}{\beta-it}\quad\mbox{and}\quad \nabla_\vartheta\zeta(t,\vartheta)=((\beta-it)^{-1},-\alpha(\beta-it)^{-2})^\top,\quad t\in\R.
 \end{equation*}
Note that
\begin{equation*}
      \overline{\nabla_\vartheta\zeta(t,\vartheta)}=((\beta+it)^{-1},-\alpha(\beta+it)^{-2})^\top=\nabla_\vartheta\zeta(-t,\vartheta),\quad t\in\R.
 \end{equation*}
The moment estimators are $\widehat{\alpha}_n=\overline{x}_n^2/s_x^2$ and $\widehat{\beta}_n=\overline{x}_n/s_x^2$. Some calculations show that
\begin{equation*}
    \ell(x,\alpha,\beta)=\left(2x-\alpha-(x-\alpha)^2,\frac{\beta}{\alpha}\left((2-\beta)x-(x-\alpha)^2\right)\right)^\top.
\end{equation*}
Some more calculations show
\begin{equation*}
    I(\vartheta)=\mathbb{E} \left[\ell(X_\infty,\vartheta)\ell(X_\infty,\vartheta)^\top\right]=\beta^{-2}\left(\begin{array}{cc}m_{11}(\alpha,\beta) & m_{12}(\alpha,\beta) \\
     m_{12}(\alpha,\beta) & m_{22}(\alpha,\beta)
\end{array}\right),
\end{equation*}
where
\begin{eqnarray*}
    m_{11}(\alpha,\beta)&=&\frac{\alpha \left( \alpha+1 \right)}{\beta^2}  \left(  \left( \beta-1 \right) ^{4}{
\alpha}^{2}+ \left( {\beta}^{2}-2\beta+5 \right)  \left( \beta-1
 \right) ^{2}\alpha+4{\beta}^{2}-8\beta+6 \right) \\
    m_{12}(\alpha,\beta)&=&- \frac{\alpha+1}{\beta}  \left(  \left( \beta-1 \right) ^{4}{\alpha}^{2}+ \left( {\beta}^{2}-2
\beta+5 \right)  \left( \beta-1 \right) ^{2}\alpha-2{\beta}^{3}+6
{\beta}^{2}-8\beta+6 \right)
,\\
    m_{22}(\alpha,\beta)&=& (\beta-1)^4 {
\alpha}^{2}+ 2 \left( {\beta}^{2}-2\beta+3 \right)  \left( \beta-1 \right) ^{2} \alpha\\&&+\left( {\beta}^{2}-2\beta+2 \right)  \left( {\beta}^{2}-6\beta+4
 \right) +3+{\frac {\left( {\beta}^{2}-2\beta+2 \right) ^{2}+2
}{\alpha}}.
\end{eqnarray*}
Writing
\begin{equation*}
    K_\vartheta^{(0)}(s,t)={\frac { \left( -{\beta}^{2}+ \left(s-t\right) i\beta-st \left(
\alpha+1 \right)  \right) \alpha}{ \left( i(s-t)-\beta \right) ^{2}
 \left( is-\beta \right)  \left( \beta+it \right) }}\varphi(s-t)
\end{equation*}
and
\begin{equation*}
    K_\vartheta^{(1)}(s,t)=\nabla_\vartheta\zeta(s,\vartheta)^\top I(\vartheta)\nabla_\vartheta\zeta(-t,\vartheta)\varphi(s)\varphi(-t).
\end{equation*}
Furthermore, we write
\begin{eqnarray*}
    K_\vartheta^{(2)}(s,t)&=&\varphi(s)\nabla_\vartheta\zeta(s,\vartheta)^\top \left(\begin{array}{c}v_1(t,\vartheta)\\v_2(t,\vartheta)\end{array}\right),
\end{eqnarray*}
where
\begin{eqnarray*}
v_1(t,\vartheta)&=&\zeta(-t,\vartheta)(\alpha+1)\left(\alpha\varphi(-t)+i(it+\beta+2)\varphi'(-t)\right)\\&&-(2\alpha+2+\zeta(-t,\vartheta))\varphi''(-t))-i\varphi'''(-t)),\\
v_2(t,\vartheta)&=&\frac{\beta}{\alpha}\bigg[\alpha^2\zeta(-t,\vartheta)\varphi(-t)-i\left((\beta-2\alpha-2)\zeta(-t,\vartheta)-\alpha^2\right)\varphi'(-t)\\
&&-(\zeta(-t,\vartheta)+2(\alpha+1)-\beta)\varphi''(-t)-i\varphi'''(-t)\bigg].
\end{eqnarray*}
With these notations the covariance kernel in Theorem \ref{thm:CLTep} is given after cumbersome long calculations by
\begin{equation*}
    K_\vartheta(s,t)=K_\vartheta^{(0)}(s,t)+K_\vartheta^{(1)}(s,t)-K_\vartheta^{(2)}(s,t)-K_\vartheta^{(2)}(-s,-t),\quad s,t\in\R.
\end{equation*}
\end{example}

\section{Simulation results}\label{sec:simuSM}

In this section we provide the simulation results for the Monte Carlo simulations involving the parametric bootstrap procedures of Section \ref{sec:Simu}. Note that we chose the weight functions $\omega_1(t)\propto \mathrm{exp}({-\gamma |t|})$ and $\omega_2(t) \propto \mathrm{exp}({-\gamma t^2})$, $\gamma>0$.

\subsection{Testing Poissonity}
In this subsection we focus on testing the fit of discrete data to the Poisson family (Example \ref{ex:constaddsizebias}, item \ref{it:1}) by applying the test statistics given in Example \ref{ex:adsibicont} for   different tuning parameters $\gamma>0$. Note that this problem has been studied in the literature, see \cite{GH:2000} for a review and Section 5 of \cite{BEN:2022} for new methods and recent references.

We simulate 38 representatives of families of distributions. In order to show that all the considered testing procedures maintain the nominal level $\upalpha$ of 10\%, we consider the Po$(\lambda)$ distribution with $\lambda\in\{1,5,10,30\}$. As alternatives we consider the discrete uniform distribution $\mathcal{U}\{0;m\}$ on the values $\{0,1,\ldots,m\}$ with $m\in\{1,2,\ldots,6,10\}$, the binomial distribution Bin$(m,{\rm p})$ with $m\in\{2,4,10,20,50\}$ and ${\rm p}\in\{0.1,0.25,0.5\}$, the negative binomial distribution $\mathcal{NB}(r,{\rm p})$, with $r\in\{1,2,3,5,9,10,15,45\}$ and ${\rm p}\in\{0.5,2/3,0.75,0.9\}$, Poisson mixtures of the form $\mathcal{PP}({\rm p};\vartheta_1,\vartheta_2)={\rm p}\mbox{Po}(\vartheta)+(1-{\rm p})\mbox{Po}(5)$ for $\vartheta\in\{1,2,3\}$ and ${\rm p}\in\{0.01,0.05,0.25,0.5\}$, a 0.9/0.1 mixture of Po(3) and point mass in 0 denoted by $\mathcal{P}(3)\delta_0$, the discrete Weibull distribution $\mathcal{W}(\vartheta_1,\vartheta_2)$ with $\vartheta_1\in\{0.1,0.25,0.5,0.75,0.9\}$ and $\vartheta_2\in\{1,2,3\}$. For pseudo random number generation of $\mathcal{U}$ and $\mathcal{W}$ the package \texttt{extraDistr}, see \cite{W:2019}, was used. Note that a major part of these alternatives is found in the simulation study presented in \cite{GH:2000}, Table 1 and 2, for comparison to other test statistics.

\begin{table}[!t]
\caption{Empirical rejection rates for the tests of poisson distribution using the size bias transform ($\upalpha=0.1$, 10000 repetitions, 500 bootstrap samples)}\label{tab:Poisson}
\centering
\begin{tabular}{ lrrrrrrrrrrrr }
 && \multicolumn{ 5 }{c}{ $T_{\omega_1}$ } &  & \multicolumn{ 5 }{c}{ $T_{\omega_2}$ } \\
Dist. / $\gamma$ &  & 0.25 & 0.5 & 1 & 3 & 5 &  & 0.25 & 0.5 & 1 & 3 & 5 \\[1mm]\hline
Po(1)                   &  & 10 & 9 & 9 & 9 & 9 &  & 9 & 9 & 9 & 9 & 9 \\
Po(5)                   &  & 10 & 10 & 10 & 9 & 9 &  & 10 & 10 & 10 & 10 & 9 \\
Po(10)                  &  & 10 & 10 & 10 & 10 & 10 &  & 9 & 10 & 10 & 9 & 10 \\
Po(30)                  &  & 10 & 11 & 10 & 10 & 10 &  & 10 & 10 & 10 & 10 & 10 \\\hline
$\mathcal{U}(0;1)$      &  & 99 & 100 & 100 & 100 & 100 &  & 99 & 99 & 100 & 100 & 100 \\
$\mathcal{U}(0;2)$      &  & 74 & 91 & 92 & 87 & 87 &  & 67 & 75 & 86 & 92 & 91 \\
$\mathcal{U}(0;4)$      &  & 63 & 82 & 82 & 73 & 72 &  & 46 & 72 & 82 & 84 & 80 \\
$\mathcal{U}(0;5)$      &  & 81 & 89 & 84 & 69 & 67 &  & 76 & 86 & 89 & 85 & 79 \\
$\mathcal{U}(0;6)$      &  & 92 & 93 & 87 & 66 & 62 &  & 91 & 93 & 93 & 88 & 80 \\
$\mathcal{U}(0;10)$     &  & 100 & 99 & 92 & 56 & 50 &  & 100 & 100 & 99 & 91 & 77 \\
Bin(2,5)              &  & 98 & 98 & 95 & 80 & 77 &  & 98 & 98 & 98 & 96 & 92 \\
Bin(4,0.25)             &  & 32 & 28 & 21 & 16 & 14 &  & 32 & 33 & 30 & 23 & 19 \\
Bin(10,0.1)             &  & 11 & 10 & 9 & 10 & 9 &  & 11 & 11 & 11 & 10 & 10 \\
Bin(10,0.5)             &  & 89 & 69 & 35 & 17 & 15 &  & 91 & 82 & 65 & 34 & 24 \\
Bin(20,0.25)            &  & 26 & 17 & 12 & 11 & 11 &  & 28 & 23 & 16 & 12 & 11 \\
Bin(50,0.1)             &  & 11 & 10 & 10 & 10 & 10 &  & 11 & 11 & 10 & 10 & 10 \\
$\mathcal{NB}(1;0.5)$   &  & 90 & 87 & 78 & 62 & 58 &  & 90 & 89 & 87 & 79 & 71 \\
$\mathcal{NB}(2;2/3)$   &  & 60 & 55 & 41 & 29 & 26 &  & 60 & 59 & 53 & 42 & 36 \\
$\mathcal{NB}(3;0.75)$  &  & 41 & 37 & 27 & 17 & 18 &  & 41 & 39 & 35 & 27 & 22 \\
$\mathcal{NB}(9;0.9)$   &  & 16 & 15 & 12 & 10 & 11 &  & 16 & 16 & 14 & 12 & 11 \\
$\mathcal{NB}(5;0.5)$   &  & 90 & 74 & 37 & 14 & 13 &  & 89 & 82 & 66 & 31 & 19 \\
$\mathcal{NB}(10;2/3)$  &  & 53 & 35 & 17 & 10 & 10 &  & 54 & 42 & 31 & 16 & 12 \\
$\mathcal{NB}(15;0.75)$ &  & 35 & 23 & 13 & 10 & 10 &  & 35 & 28 & 20 & 13 & 11 \\
$\mathcal{NB}(45;0.9)$  &  & 14 & 12 & 10 & 9 & 9 &  & 15 & 13 & 12 & 10 & 10 \\
$\mathcal{PP}(0.5;1,5)$ &  & 100 & 100 & 95 & 63 & 55 &  & 100 & 100 & 99 & 94 & 83 \\
$\mathcal{PP}(0.5;2,5)$ &  & 79 & 67 & 38 & 17 & 15 &  & 80 & 75 & 63 & 36 & 24 \\
$\mathcal{PP}(0.5;3,5)$ &  & 29 & 20 & 14 & 10 & 10 &  & 29 & 24 & 19 & 13 & 11 \\
$\mathcal{PP}(0.25;1,5)$ &  & 91 & 82 & 54 & 22 & 20 &  & 92 & 88 & 79 & 52 & 34 \\
$\mathcal{PP}(0.05;1,5)$ &  & 20 & 15 & 11 & 10 & 10 &  & 20 & 18 & 15 & 11 & 11 \\
$\mathcal{PP}(0.01;1,5)$ &  & 11 & 10 & 9 & 10 & 10 &  & 11 & 10 & 11 & 9 & 9 \\
$\mathcal{P}\delta_0(9;3)$ &  & 43 & 37 & 26 & 17 & 16 &  & 42 & 41 & 36 & 26 & 22 \\
$\mathcal{W}(0.5,1)$ &  & 90 & 87 & 79 & 61 & 58 &  & 90 & 89 & 87 & 78 & 71 \\
$\mathcal{W}(0.25,1)$ &  & 39 & 38 & 34 & 30 & 28 &  & 39 & 39 & 39 & 35 & 32 \\
$\mathcal{W}(0.75,1)$ &  & 100 & 100 & 99 & 86 & 81 &  & 100 & 100 & 100 & 99 & 96 \\
$\mathcal{W}(0.5,2)$ &  & 59 & 58 & 53 & 50 & 49 &  & 59 & 58 & 58 & 54 & 54 \\
$\mathcal{W}(0.25,2)$ &  & 14 & 15 & 16 & 16 & 18 &  & 13 & 13 & 14 & 15 & 17 \\
$\mathcal{W}(0.75,1)$ &  & 43 & 37 & 27 & 19 & 18 &  & 44 & 43 & 38 & 29 & 24 \\
$\mathcal{W}(0.1,1)$ &  & 14 & 14 & 16 & 16 & 15 &  & 14 & 14 & 14 & 14 & 16 
\end{tabular}%
\end{table}

\subsection{Testing for gamma distribution}\label{subsec:GDSM}

As alternative families of distributions we consider the following distributions with scale parameter fixed to 1 (which can be done w.l.o.g. due to invariance properties of the considered estimators): The Weibull distribution $W(\vartheta)$, the inverse Gaussian law $IG(\vartheta)$, the lognormal law $LN(\vartheta)$, the power distribution $PW(\vartheta)$, the shifted-Pareto distribution $SP(\vartheta)$, the Gompertz law $GO(\vartheta)$ and the linear increasing failure rate law $LF(\vartheta)$. For details on the densities of these probability laws, see \cite{HME:2012}. We chose these families, $n=50$ and a significance level of 0.05 for easy comparison to the existing simulation studies in \cite{BE:2019}. 

\begin{table}[!t]
\caption{Empirical rejection rates for the tests of gamma distribution using the size bias transform ($\upalpha=0.05$, 10000 repetitions, 500 bootstrap samples)}\label{tab:Gamma}
\centering
\begin{tabular}{ lrrrrrrrrrr }
   && \multicolumn{ 4 }{c}{ $T_{\omega_1}$ } &  & \multicolumn{ 4 }{c}{ $T_{\omega_2}$ } \\
Dist. / $\gamma$ &  & 1 & 3 & 5 & 10 &  & 1 & 3 & 5 & 10 \\[1mm]\hline
$\Gamma(0.25)$ &  &  4 &  4 &  5 &  4 &  &  4 &  4 &  4 &  4 \\
$\Gamma(0.5)$ &  &  3 &  4 &  4 &  5 &  &  4 &  4 &  4 &  4 \\
$\Gamma(1)$ &  &  2 &  3 &  4 &  4 &  &  3 &  4 &  4 &  4 \\
$\Gamma(5)$ &  &  1 &  1 &  2 &  3 &  &  1 &  2 &  2 &  3 \\
$\Gamma(10)$ &  &  0 &  1 &  2 &  2 &  &  1 &  2 &  2 &  2 \\\hline
W$(0.5)$ &  & 22 & 31 & 32 & 32 &  & 30 & 33 & 33 & 33 \\
W$(1.5)$ &  &  1 &  3 &  3 &  4 &  &  3 &  3 &  3 &  3 \\
W$(3)$ &  &  1 &  8 & 14 & 18 &  &  5 & 12 & 15 & 17 \\
IG(0.5) &  & 32 & 66 & 73 & 74 &  & 57 & 70 & 73 & 73 \\
IG(1.5) &  &  5 & 28 & 38 & 43 &  & 18 & 34 & 38 & 43 \\
IG(3) &  &  1 & 11 & 19 & 26 &  &  7 & 15 & 19 & 24 \\
LN(0.5) &  &  1 &  9 & 17 & 24 &  &  5 & 13 & 17 & 22 \\
LN(0.8) &  &  8 & 33 & 42 & 48 &  & 22 & 39 & 44 & 47 \\
LN(1.5) &  & 53 & 75 & 75 & 77 &  & 69 & 76 & 77 & 77 \\
PW(1) &  & 64 & 91 & 91 & 86 &  & 90 & 90 & 88 & 86 \\
PW(2) &  & 52 & 62 & 53 & 51 &  & 64 & 52 & 50 & 50 \\
PW(4) &  & 20 & 12 & 12 & 16 &  & 12 & 10 & 12 & 15 \\
SP(1) &  & 79 & 88 & 89 & 89 &  & 86 & 89 & 89 & 89 \\
SP(2) &  & 38 & 57 & 60 & 60 &  & 52 & 60 & 60 & 60 \\
Go(2) &  &  9 & 36 & 42 & 46 &  & 33 & 40 & 42 & 44 \\
Go(4) &  & 15 & 58 & 65 & 70 &  & 51 & 64 & 66 & 68 \\
LF(2) &  &  3 &  8 &  9 & 12 &  &  7 &  9 & 10 & 11 \\
LF(4) &  &  3 & 11 & 13 & 16 &  &  9 & 12 & 14 & 15
\end{tabular}
\end{table}

\subsection{Testing for the generalized Dickman distribution}
We borrow the alternatives from Subsection \ref{subsec:GDSM} and show the results in Table \ref{tab:Dickman}. Here we fix the nominal level to $\upalpha=0.1$. 

\begin{table}[!t]
\caption{Empirical rejection rates for the new tests of Dickman distribution using the size bias transform ($\upalpha=0.1$, 10000 repetitions, 500 bootstrap samples)}\label{tab:Dickman}
\centering
\begin{tabular}{ lrrrrrrrrrrrr }
  && \multicolumn{ 5 }{c}{ $T_{\omega_1}$ } &  & \multicolumn{ 5 }{c}{ $T_{\omega_2}$ } \\
Dist. / $\gamma$ &  & 0.25 & 0.5 & 1 & 3 & 5 &  & 0.25 & 0.5 & 1 & 3 & 5 \\[1mm]\hline
$\mathcal{D}(1)$ &  & 8 & 9 & 10 & 9 & 10 &  & 9 & 9 & 9 & 10 & 10 \\
$\mathcal{D}(2)$ &  & 8 & 9 & 9 & 9 & 9 &  & 10 & 10 & 10 & 10 & 10 \\
$\mathcal{D}(3)$ &  & 9 & 9 & 10 & 10 & 10 &  & 9 & 10 & 10 & 10 & 10 \\
$\mathcal{D}(5)$ &  & 8 & 10 & 9 & 10 & 10 &  & 10 & 10 & 10 & 10 & 10 \\
$\mathcal{D}(10)$ &  & 9 & 10 & 10 & 9 & 11 &  & 10 & 10 & 9 & 10 & 10 \\\hline
$\Gamma(0.25,1)$ &  &  45 &  53 &  64 &  69 &  70 &  &  65 &  69 &  70 &  70 &  69 \\ 
$\Gamma(0.5,2)$ &  &  14 &  31 &  38 &  36 &  33 &  &  39 &  37 &  35 &  32 &  33 \\ 
$\Gamma(1,3)$ &  &  15 &  47 &  61 &  63 &  62 &  &  64 &  64 &  63 &  61 &  61 \\ 
$\Gamma(5,4)$ &  &   9 &  27 &  57 &  90 &  94 &  &  66 &  82 &  90 &  94 &  94 \\ 
$\Gamma(10,5)$ &  &  10 &  27 &  60 &  98 &  99 &  &  71 &  91 &  98 & 100 & 100 \\ 
W(0.5) &  & 100 & 100 & 100 & 100 & 100 &  & 100 & 100 & 100 & 100 & 100 \\ 
W(1.5) &  &   6 &  10 &  13 &  18 &  21 &  &  15 &  17 &  20 &  21 &  21 \\ 
W(3) &  &  66 &  99 & 100 & 100 & 100 &  & 100 & 100 & 100 & 100 & 100 \\ 
IG(0.5) &  &  78 &  83 &  92 &  97 &  97 &  &  93 &  96 &  97 &  97 &  96 \\ 
IG(1.5) &  &  21 &  37 &  52 &  63 &  57 &  &  56 &  62 &  63 &  55 &  50 \\ 
IG(3) &  &  15 &  45 &  68 &  81 &  76 &  &  72 &  78 &  80 &  75 &  70 \\ 
LN(0.5) &  &  14 &  43 &  68 &  85 &  84 &  &  74 &  82 &  84 &  80 &  78 \\ 
LN(0.8) &  &  37 &  40 &  56 &  85 &  87 &  &  57 &  72 &  82 &  87 &  86 \\ 
LN(1.5) &  &  99 & 100 & 100 & 100 & 100 &  & 100 & 100 & 100 & 100 & 100 \\ 
PW(1) &  &  93 & 100 & 100 & 100 & 100 &  & 100 & 100 & 100 & 100 & 100 \\ 
PW(2) &  &  21 &  60 &  86 &  96 &  97 &  &  92 &  96 &  96 &  97 &  97 \\ 
PW(4) &  &   5 &   9 &  17 &  28 &  31 &  &  20 &  25 &  29 &  31 &  32 \\ 
SP(1) &  & 100 & 100 & 100 & 100 & 100 &  & 100 & 100 & 100 & 100 & 100 \\ 
SP(2) &  &  84 &  86 &  92 &  97 &  97 &  &  93 &  96 &  97 &  98 &  97 \\ 
Go(2) &  &   7 &  11 &  17 &  35 &  39 &  &  20 &  27 &  35 &  39 &  40 \\ 
Go(4) &  &   9 &  15 &  25 &  36 &  28 &  &  28 &  36 &  35 &  27 &  24 \\ 
LF(2) &  &   7 &  26 &  50 &  76 &  80 &  &  59 &  69 &  76 &  80 &  82 \\ 
LF(4) &  &  27 &  79 &  96 &  99 & 100 &  &  98 &  99 &  99 & 100 & 100  
\end{tabular}
\end{table}

\subsection{Testing the fit to the family of compound Poisson exponential distributions}

The compound Poisson Exponential distribution is denoted by CP$(\lambda,\mbox{Exp}(\beta))$, where $\lambda>0$ is the parameter of the Poisson distribution and $\beta>0$ the parameter of the exponential law, see Example \ref{ex:compoupoi} item 1. We borrow the alternatives from Subsection \ref{subsec:GDSM} and add a mixed compound Poisson exponential model by simulating MCP$(\lambda_1,\beta_1,\lambda_2,\beta_2;{\rm p})={\rm p}\,\mbox{CP}(\lambda_1,\mbox{Exp}(\beta_1))+(1-{\rm p})\mbox{CP}(\lambda_2,\mbox{Exp}(\beta_2))$ for $\lambda_1=\beta_1=1$ and $\lambda_2\in\{1,2,3,4,5,10,20,50\}$, $\beta_2\in\{5,6\}$, and ${\rm p}\in\{0.25,0.5,0.75\}$. Here we fixed the nominal level to $\upalpha=0.05$. We show the results in Table \ref{tab:CPE}. Note that the procedure is conservative in view of the type-I error and shows acceptable performance against most of the alternatives, although also practically blind against some alternatives. Note that the sample size is not large, so for bigger data sets the power performance will increase. It would be interesting to compare the power performance to the methods of \cite{GJM:2022} and \cite{BG:2023}.

\begin{table}[!t]
\caption{Empirical rejection rates for the tests $T_{\omega_3}$ with $\omega_3(t)\propto \exp(-\gamma t^2/2)$ of compound Poisson exponential distribution using the size bias transform ($n=50$, $\upalpha=0.05$, 10000 repetitions, 500 bootstrap samples)}\label{tab:CPE}
\centering
\begin{tabular}{ lrrrrr }
Dist. / $\gamma$ &  & 1 & 3 & 5 & 10 \\[1mm]\hline
CP(1,Exp(1)) &  & 2 & 2 & 3 & 3 \\
CP(2,Exp(2)) &  & 3 & 4 & 4 & 3 \\
CP(3,Exp(2)) &  & 2 & 3 & 3 & 3 \\
CP(4,Exp(2)) &  & 2 & 3 & 3 & 4 \\ \hline
$\Gamma(0.25)$ &  & 5 & 7 & 5 & 4 \\
$\Gamma(0.5)$ &  & 7 & 10 & 10 & 9 \\
$\Gamma(1)$ &  & 5 & 9 & 10 & 10 \\
$\Gamma(5)$ &  & 2 & 3 & 4 & 6 \\
$\Gamma(10)$ &  & 1 & 2 & 2 & 3 \\
IG(0.5) &  & 19 & 35 & 40 & 42 \\
IG(1.5) &  & 34 & 40 & 40 & 37 \\
IG(3) &  & 31 & 31 & 31 & 30 \\
LN(0.5) &  & 32 & 33 & 32 & 31 \\
LN(0.8) &  & 25 & 42 & 45 & 46 \\
LN(1.5) &  & 1 & 2 & 3 & 6 \\
PW(1) &  & 80 & 78 & 77 & 74 \\
PW(2) &  & 32 & 29 & 26 & 24 \\
PW(4) &  & 5 & 4 & 3 & 3 \\
SP(1) &  & 0 & 0 & 1 & 2 \\
SP(2) &  & 9 & 22 & 29 & 35 \\
Go(2) &  & 17 & 16 & 15 & 14 \\
Go(4) &  & 34 & 39 & 38 & 37 \\
W$(3,2)$ &  & 10 & 12 & 12 & 12 \\
W$(2,5)$ &  & 2 & 3 & 3 & 3 \\
MCP(1,1,1,6;0.25) &  & 11 & 12 & 13 & 10 \\
MCP(1,1,2,6;0.25) &  & 30 & 34 & 34 & 30 \\
MCP(1,1,3,6;0.25)&  & 36 & 40 & 38 & 39 \\
MCP(1,1,4,6;0.25) &  & 29 & 37 & 37 & 39 \\
MCP(1,1,1,5;0.5) &  & 3 & 2 & 5 & 4 \\
MCP(1,1,2,5;0.5) &  & 5 & 11 & 11 & 10 \\
MCP(1,1,3,5;0.5) &  & 08 & 15 & 16 & 15 \\
MCP(1,1,4,5;0.5) &  & 10 & 15 & 17 & 18 \\
MCP(1,1,5,5;0.75) &  & 2 & 5 & 7 & 6 \\
MCP(1,1,10,5;0.75) &  & 7 & 8 & 10 & 10 \\
MCP(1,1,20,5;0.75) &  & 13 & 29 & 31 & 31 \\
MCP(1,1,50,5;0.75) &  & 09 & 29 & 57 & 81 \\
\end{tabular}%
\end{table}

\end{document}